\documentclass[aip,jcp,reprint,groupedaddress]{revtex4-2}%
\usepackage{graphicx}
\usepackage{tabularx}
\usepackage{bm}
\usepackage{amsmath}
\usepackage{amsfonts}
\usepackage{amssymb}
\usepackage{float}%
\providecommand{\U}[1]{\protect\rule{.1in}{.1in}}
\begin{document}
\title{Thawed Gaussian Ehrenfest dynamics}
\author{Ji\v{r}\'i Van\'i\v{c}ek}
\email{jiri.vanicek@epfl.ch}
\affiliation{Laboratory of Theoretical Physical Chemistry, Institut des Sciences et
Ing\'enierie Chimiques, Ecole Polytechnique F\'ed\'erale de Lausanne (EPFL),
CH-1015, Lausanne, Switzerland}
\date{\today}

\begin{abstract}
Ehrenfest dynamics is a widely used mixed quantum--classical approach for
nonadiabatic molecular dynamics, whereas thawed Gaussian wavepacket dynamics
provides an efficient semiclassical description of adiabatic nuclear quantum
dynamics. Here we describe thawed Gaussian Ehrenfest dynamics (TGED), which
unifies and generalizes these two methods to capture both electronic
nonadiabaticity and nuclear quantum effects within a single framework. The
fully variational formulation of TGED is derived by applying the
time-dependent variational principle to a Hartree product of electronic and
Gaussian nuclear wavepackets. Replacing the effective locally quadratic
molecular potential obtained from this variational treatment by alternative
effective locally quadratic potentials yields an infinite family of TGED
methods, of which we present several members. We analyze the limiting cases of
the general formalism and show, in particular, that it reduces to conventional
Ehrenfest dynamics in the classical limit for the nuclei and to thawed
Gaussian wavepacket dynamics in the absence of electronic coupling. Finally,
we present explicit geometric integrators for the entire family of methods and
identify the conditions under which the different approximations become exact.

\end{abstract}
\maketitle


\graphicspath{
{./figures/}{C:/Users/Jiri/Dropbox/Papers/Chemistry_papers/2026/Ehrenfest_TGA/figures/}}

\section{Introduction}

Due to their vastly different masses, nuclei and electrons in molecules evolve
on very different time scales, which often permits their separation within the
Born--Oppenheimer approximation.\cite{Born_Oppenheimer:1927,book_Tannor:2007}
This approximation, however, breaks down near conical intersections and, more
generally, whenever adiabatic potential-energy surfaces become nearly
degenerate.\cite{Domcke_Yarkony:2012} In such situations, an accurate
description of molecular quantum dynamics requires nonadiabatic simulations
that explicitly account for the coupling between electronic and nuclear
motion.\cite{Agostini_Curchod:2019}

Because exact quantum simulations in either the
adiabatic\cite{Choi_Vanicek:2019} or diabatic\cite{Roulet_Vanicek:2019}
electronic representation, as well as exact-factorization
approaches,\cite{Abedi_Gross:2010} remain limited to systems with only a few
degrees of freedom, more efficient low-rank-tensor-based methods have been
developed, most notably the multiconfigurational time-dependent Hartree
(MCTDH) method.\cite{Meyer_Cederbaum:1990} Although such methods substantially
reduce the exponential scaling with system size, they remain prohibitively
expensive for on-the-fly \textit{ab initio} simulations. This limitation has
motivated the development of multi-trajectory Gaussian-basis methods,
including multiple spawning,\cite{Martinez_Levine:1997} variational
multiconfigurational Gaussians,\cite{Worth_Burghardt:2004} and
multiconfigurational Ehrenfest dynamics.\cite{Shalashilin:2009} While these
approaches can, in principle, converge to the exact quantum solution,
practical convergence typically requires a large number of trajectories and is
complicated by numerical difficulties associated with the nonorthogonality of
Gaussian basis functions.

For greater computational efficiency, albeit at the expense of accuracy, many
researchers have turned to mixed quantum--classical
methods,\cite{Kapral_Ciccotti:1999,Huo_Coker:2011,Runeson_Richardson:2020,Wu_Liu:2025}
which treat electrons quantum mechanically while describing nuclei
classically. On the one hand, numerous
extensions\cite{Belyaev_Trigila:2014,Subotnik_Bellonzi:2016,Mannouch_Richardson:2023}
of Tully's fewest-switches surface-hopping method\cite{Tully:1990} were
developed to describe nonadiabatic transitions between electronic states as
well as the correlation between the nuclear and electronic degrees of freedom.
On the other hand, there exist various mean-field approaches, including the
single-trajectory Ehrenfest dynamics,\cite{Ehrenfest:1927} and locally
mean-field methods, such as the multi-trajectory Ehrenfest
dynamics.\cite{Tully:1998} More recently, the single-potential-evaluation
Ehrenfest dynamics (SPEED) was introduced to reduce the computational cost of
multi-trajectory Ehrenfest dynamics to that of propagating a single \textit{ab
initio} trajectory while retaining some non-mean-field
effects.\cite{Scheidegger_Vanicek:2025a} Despite its mean-field nature,
Ehrenfest dynamics has proved useful in many
applications\cite{Zeiri_Kosloff:1990,Akimov_Prezhdo:2014,Gherib_Izmaylov:2015}
because it captures some nonadiabatic effects and is known to be exact under
well-defined conditions (see Sec.~\ref{sec:exactness_ED}).

A major limitation of mixed quantum--classical methods, including both surface
hopping and Ehrenfest dynamics, is that they neglect nuclear quantum effects
altogether. Such effects can instead be approximately described using
semiclassical approximations, including initial-value
representation\cite{Miller:2001} of the Van Vleck propagator,\cite{Vleck:1928}
the Herman-Kluk propagator\cite{Herman_Kluk:1984} and their
extensions.\cite{Ceotto_Conte:2017} Among semiclassical approaches, the
simplest is Heller's single-trajectory thawed Gaussian wavepacket dynamics
(TGWD)\cite{Heller:1975,book_Heller:2018} and its
variants.\cite{Coalson_Karplus:1990,Begusic_Vanicek:2019,Vanicek:2023,Burkhard_Lasser:2024}
TGWD incorporates nuclear quantum effects while accounting, at least
approximately, for anharmonicity. Combined with on-the-fly \textit{ab initio}
electronic-structure calculations, it has been successfully applied to
vibrationally resolved electronic spectroscopy at both
zero\cite{Wehrle_Vanicek:2014,Kletnieks_Vanicek:2023} and
finite\cite{Begusic_Vanicek:2020,Begusic_Vanicek:2021} temperatures, as well
as to studies of electronic coherence and
decoherence.\cite{Golubev_Vanicek:2020,Scheidegger_Golubev:2022,Scheidegger_Vanicek:2025}
Moreover, TGWD is exact for globally harmonic potentials
(Sec.~\ref{sec:exactness_TGWD}).

\begin{figure}
[H]
\centering\includegraphics[width=\textwidth]{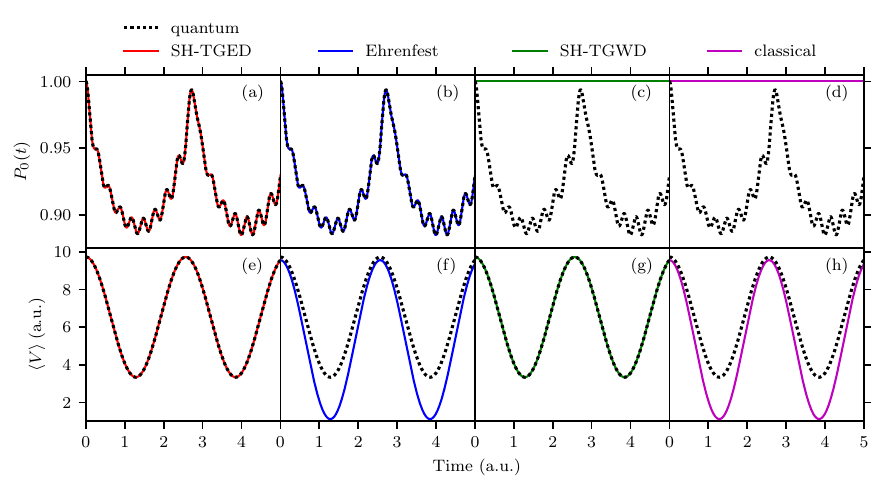}\caption{{Comparison
of thawed Gaussian Ehrenfest dynamics (TGED) with Ehrenfest dynamics, thawed
Gaussian wavepacket dynamics (TGWD), and purely classical dynamics in a
system of ten vertically displaced two-dimensional harmonic oscillators
coupled by a coordinate-independent electronic coupling. The exact quantum results
are indicatd by the dashed line. The upper panels (a)--(d)
quantify the nonadiabatic dynamics by showing the time-dependent population $P_{0}(t)$ of
the ground electronic state, whereas the lower panels (e)--(h) illustrate nuclear
quantum effects through the expectation value $\langle V \rangle$ of the potential energy. For
this model system, TGED is exact (see Sec.~\ref{sec:exactness_TGED}). [The
single-Hessian (SH) TGED from Sec.~\ref{sec:SH-TGED} was used, but all TGED
variants from Sec.~\ref{sec:TGED_family} would be exact here.]
Ehrenfest dynamics reproduces the nonadiabatic dynamics exactly but fails to
capture nuclear quantum effects. Conversely, TGWD accurately describes the
nuclear quantum dynamics but cannot account for nonadiabatic transitions.
Purely classical dynamics captures neither nonadiabatic nor nuclear quantum
effects.}}\label{fig:TGED_comparison}
\end{figure}

Motivated by the complementary strengths of Ehrenfest dynamics and TGWD, here
we describe \textquotedblleft thawed Gaussian Ehrenfest
dynamics\textquotedblright\ (TGED), which unifies and generalizes both
approaches. Whereas Ehrenfest dynamics captures nonadiabatic effects but
neglects nuclear quantum effects, TGWD includes nuclear quantum effects but is
restricted to a single potential energy surface. TGED combines these two
complementary descriptions within a single framework. Although the resulting
method remains a mean-field approximation and is therefore necessarily
approximate, it captures both nonadiabatic and nuclear quantum effects (see
Fig.~\ref{fig:TGED_comparison}) while retaining the computational efficiency
of single-trajectory propagation. Whereas the severe limitations of mean-field
methods are well known, Fig.~\ref{fig:TGED_comparison} intentionally shows the
results for a system, in which the mean-field TGED is exact, but neither TGWD
nor Ehrenfest dynamics is. The system, which consists of ten vertically
displaced two-dimensional harmonic potentials coupled with constant couplings,
is simple because the dynamics of nuclei and electrons is uncorrelated, yet
there exist realistic systems whose Hamiltonians are not very different. The
TGED\ method possesses well-defined limiting cases
(Sec.~\ref{sec:limits_of_TGED}) and becomes exact for clearly identifiable
classes of Hamiltonians (Sec.~\ref{sec:exactness_TGED}). The strengths and
limitations of TGED in applications to nonadiabatic dynamics in the vicinity
of conical intersections will be explored
elsewhere.\cite{Scheidegger_Vanicek:2026}

The remainder of this paper is organized as follows.
Section~\ref{sec:background} reviews the time-dependent Hartree approximation
for the molecular wavefunction and specializes it to the diabatic
representation and separable Hamiltonians. In Sec.~\ref{sec:TGED}, we derive
the general thawed Gaussian Ehrenfest dynamics by combining the time-dependent
Hartree approximation with a Gaussian ansatz for the nuclear wavepacket.
Section~\ref{sec:TGED_family} introduces an infinite family of TGED methods
that differ in the choice of the effective local quadratic potential-energy
matrix. In Sec.~\ref{sec:limits_of_TGED}, we show how different limits of the
TGED recover Gaussian wavepacket dynamics, Ehrenfest dynamics, or classical
nuclear dynamics. Section~\ref{sec:integrators} presents geometric integrators
for the general TGED equations of motion, while Sec.~\ref{sec:exactness}
establishes the conditions under which the various approximations become
exact. Finally, Sec.~\ref{sec:conclusion} concludes the paper.

\onecolumngrid

\twocolumngrid

\section{Theoretical background\label{sec:background}}

Let us recall the time-dependent Hartree (TDH)\ approximation for the
molecular wavefunction, because its mixed quantum-semiclassical treatment will
yield the TGED in the next section.

\subsection{Molecular Schr\"{o}dinger equation}

Quantum evolution of a molecule is governed by the time-dependent
Schr\"{o}dinger equation (TDSE)%
\begin{equation}
i\hbar\dot{\Psi}(t)=\mathcal{H}\Psi(t), \label{eq:TDSE_mol}%
\end{equation}
where $\Psi(t)$ denotes the molecular state at time $t$ and $\mathcal{H}$ is
the molecular Hamiltonian. (In general, operators acting on both nuclei and
electrons will be denoted by a calligraphic font, whereas operators acting
only on nuclei or only on electrons will have a hat~$~\hat{}$.) It will be
convenient to express the molecular Hamiltonian as the sum%
\begin{equation}
\mathcal{H}=\mathcal{H}(\hat{q},\hat{p})=T(\hat{p})+\mathcal{V}(\hat{q})
\label{eq:H_mol}%
\end{equation}
of the nuclear kinetic energy operator $T(\hat{p})$ and the \textquotedblleft
remainder\textquotedblright\ $\mathcal{V}$, which includes the electronic
kinetic energy as well as the potential energy due to electron-electron
repulsion, nucleus-electron attraction, and nucleus-nucleus repulsion.
Although $\mathcal{V}$ is sometimes called the \textquotedblleft electronic
Hamiltonian,\textquotedblright\ the calligraphic font is used because
$\mathcal{V}$ still acts on both nuclear and electronic degrees of freedom. We
will assume that the nuclear kinetic energy is a quadratic form
\begin{equation}
T(p)=p^{T}\cdot m^{-1}\cdot p/2 \label{eq:T_p}%
\end{equation}
of the nuclear momentum $p$. In a system with $D$ nuclear degrees of freedom,
the nuclear position $q$ and momentum $p$ are $D$-dimensional vectors, whereas
the mass $m$ can, in general, be a real symmetric $D\times D$ matrix.

\subsection{Time-dependent Hartree approximation}

The TDH\ approximation\cite{Dirac:1930,book_Frenkel:1934,book_Lubich:2008}
provides the best approximate solution of the molecular TDSE
(\ref{eq:TDSE_mol}) among those, in which the molecular state can be written
as the\ \emph{Hartree product}%
\begin{equation}
\Psi(t)=a(t)\psi(t)\varphi(t) \label{eq:TDH_mol_ansatz}%
\end{equation}
of the nuclear wavepacket $\psi(t)$ and electronic wavepacket $\varphi(t)$.
The complex number $a(t)$ is inserted for convenience. We further assume that
the initial molecular state is normalized, i.e., $\left\Vert \Psi
(0)\right\Vert =1$, that $a(0)=1$, and that nuclear and electronic states are
normalized at all times $t$, $\left\Vert \psi(t)\right\Vert =\left\Vert
\varphi(t)\right\Vert =1$.

The optimal solution is found by applying the time-dependent variational
principle (TDVP)\cite{Dirac:1930,book_Frenkel:1934,book_Lubich:2008} to the
ansatz (\ref{eq:TDH_mol_ansatz}). The TDVP requires that an arbitrary
variation $\delta\Psi$ of the solution $\Psi$ satisfy the relation%
\begin{equation}
\langle\delta\Psi|i\hbar\frac{d}{dt}-\mathcal{H}|\Psi\rangle=0.
\label{eq:TDH_TDVP}%
\end{equation}
The solution, called the TDH
approximation,\cite{Dirac:1930,book_Frenkel:1934,book_Lubich:2008} conserves
both the total energy%
\begin{equation}
E\equiv\langle\mathcal{H}\rangle_{\Psi(t)}=\langle\Psi(t)|\mathcal{H}%
|\Psi(t)\rangle=\operatorname{const} \label{eq:TDH_E_conservation}%
\end{equation}
and the norm $\left\Vert \Psi(t)\right\Vert =1$ of the molecular wavefunction.
Whereas both conservation laws are enforced by the
TDVP,\cite{Lasser_Lubich:2020} energy conservation obviously requires
Hamiltonian $\mathcal{H}$ to be time-independent and norm conservation is
guaranteed by including prefactor $a(t)$ in the
ansatz~(\ref{eq:TDH_mol_ansatz}).\cite{book_Lubich:2008} This prefactor
evolves as%
\begin{equation}
a(t)=e^{iEt/\hbar}, \label{eq:TDH_a_t}%
\end{equation}
while the nuclear and electronic states satisfy the system%
\begin{align}
i\hbar\dot{\psi}  &  =\hat{H}_{n}\psi,\label{eq:TDH_mean_field_TDSE_n}\\
i\hbar\dot{\varphi}  &  =\hat{H}_{e}\varphi\label{eq:TDH_mean_field_TDSE_e}%
\end{align}
of coupled nonlinear Schr\"{o}dinger equations with mean-field nuclear and
electronic Hamiltonian operators
\begin{align}
\hat{H}_{n}  &  :=\langle\mathcal{H}\rangle_{e}\equiv\langle\varphi
(t)|\mathcal{H}|\varphi(t)\rangle,\label{eq:TDH_mean_field_H_n}\\
\hat{H}_{e}  &  :=\langle\mathcal{H}\rangle_{n}\equiv\langle\psi
(t)|\mathcal{H}|\psi(t)\rangle, \label{eq:TDH_mean_field_H_e}%
\end{align}
where $\langle\cdot\rangle_{e}:=\langle\varphi|\cdot|\varphi\rangle$ and
$\langle\cdot\rangle_{n}:=\langle\psi|\cdot|\psi\rangle$ denote the averages
over the electronic and nuclear states, respectively. These mean-field
operators satisfy the obvious identity%
\begin{equation}
\langle\hat{H}_{e}\rangle_{e}=\langle\hat{H}_{n} \rangle_{n}=E
\end{equation}
and---in contrast to the standard TDSE---the differential equations for $\psi$
and $\varphi$ are (i) coupled and (ii) nonlinear, due to the dependence of the
mean-field Hamiltonians $\hat{H}_{n}$ and $\hat{H}_{e}$ on the state of the
other subsystem. The solution expressed by Eqs.~(\ref{eq:TDH_a_t}%
)-(\ref{eq:TDH_mean_field_TDSE_e}) is unique except for an obvious gauge
freedom in distributing the phases among the three factors $a$, $\psi$, and
$\varphi$.

\subsection{Diabatic representation of the electronic state}

To express the TDH\ approximation in the diabatic representation, let us
expand the electronic state $\varphi(t)$ in an orthonormal diabatic basis
$|k\rangle$ as%
\begin{equation}
|\varphi(t)\rangle=\sum_{k}c_{k}(t)|k\rangle, \label{eq:phi_in_basis}%
\end{equation}
where $\langle j|k\rangle=\delta_{jk}$. The basis is called \textquotedblleft
diabatic\textquotedblright\ because the electronic basis states $|k\rangle$
are assumed to be independent of nuclear coordinates $q$. More precisely,
electronic states $|k\rangle$ may depend on nuclear coordinates $q$ weakly,
but this $q$-dependence can be neglected in the action of the nuclear kinetic
energy operator on the electronic
states.\cite{Choi_Vanicek:2020,Choi_Vanicek:2021}

In the diabatic basis, the time-dependent Hartree
Eqs.~(\ref{eq:TDH_mean_field_TDSE_n}) and (\ref{eq:TDH_mean_field_TDSE_e}) are
equivalent to the equations%
\begin{align}
i\hbar\dot{\psi}  &  =H_{n}(\hat{q},\hat{p};\mathbf{c})\psi
,\label{eq:TDH_TDSE_diabatic_nu}\\
i\hbar\dot{\mathbf{c}}  &  =\mathbf{H}_{e}(\psi)\mathbf{c},
\label{eq:TDH_TDSE_diabatic_el}%
\end{align}
where the mean-field nuclear Hamiltonian
\begin{equation}
H_{n}(\hat{q},\hat{p};\mathbf{c}):=\mathbf{c}^{\dag}\mathbf{H}(\hat{q},\hat
{p})\mathbf{c}%
\end{equation}
is the average over the electronic state $\mathbf{c}$ of the matrix operator
$\mathbf{H}(\hat{q},\hat{p})$, which consists of matrix elements
\begin{equation}
H_{jk}(\hat{q},\hat{p}):=\langle j|\mathcal{H}(\hat{q},\hat{p})|k\rangle
\end{equation}
of the molecular Hamiltonian $\mathcal{H}(\hat{q},\hat{p})$ and where the
electronic mean-field matrix Hamiltonian
\begin{equation}
\mathbf{H}_{e}\left(  \psi\right)  :=\langle\psi|\mathbf{H}(\hat{q},\hat
{p})|\psi\rangle\label{eq:TDH_H_e_diab}%
\end{equation}
is the average of $\mathbf{H}(\hat{q},\hat{p})$ over the nuclear state $\psi$.

\subsection{Separable Hamiltonian}

Until now, we have not taken advantage of the separable form (\ref{eq:H_mol})
of the molecular Hamiltonian $\mathcal{H}$ in the diabatic representation.
Since $\mathcal{H}$ is a sum~(\ref{eq:H_mol}) of a function of nuclear momenta
and a function of nuclear coordinates, the electronic matrix representation of
$\mathcal{H}$ is%
\begin{equation}
\mathbf{H}(\hat{q},\hat{p})=T(\hat{p})\mathbf{1}+\mathbf{V}(\hat{q}),
\end{equation}
where $\mathbf{V}(\hat{q})$ is composed of electronic matrix elements
\begin{equation}
V_{jk}(\hat{q}):=\langle j|\mathcal{V}(\hat{q})|k\rangle
\label{eq:V_matrix_els}%
\end{equation}
of the diabatic potential energy operator. The nuclear mean-field Hamiltonian
(\ref{eq:TDH_mean_field_H_n}) then becomes%
\begin{equation}
\hat{H}_{n}=T(\hat{p})+V_{n}(\hat{q};\mathbf{c}), \label{eq:H_n_in_diab_basis}%
\end{equation}
with the mean-field nuclear potential%
\begin{equation}
V_{n}(\hat{q};\mathbf{c}):=\mathbf{c}^{\dag}\mathbf{V}(\hat{q})\mathbf{c,}%
\end{equation}
while the electronic mean-field Hamiltonian, expressed in the electronic
basis, is the matrix
\begin{equation}
\mathbf{H}_{e}(\psi)=\langle T(\hat{p})\rangle_{n}\mathbf{1}+\langle
\mathbf{V}(\hat{q})\rangle_{n}.
\end{equation}
As a result, the coupled Schr\"{o}dinger equations
(\ref{eq:TDH_TDSE_diabatic_nu}) and (\ref{eq:TDH_TDSE_diabatic_el}) may be
written as
\begin{align}
i\hbar\dot{\psi}  &  =\left[  T(\hat{p})+V_{n}(\hat{q};\mathbf{c})\right]
\psi,\label{eq:TDH_TDSE_nu_separ}\\
i\hbar\dot{\mathbf{c}}  &  =[\langle T(\hat{p})\rangle_{n}+\langle
\mathbf{V}(\hat{q})\rangle_{n}]\mathbf{c}. \label{eq:TDH_TDSE_el_separ}%
\end{align}

Even if only a few electronic states are involved in the dynamics,
implementation of the mean-field TDH for a general nuclear wavefunction has a
high computational cost because it scales exponentially with $D$. As we will
now see, in the TGED, the nuclear wavefunction is a Gaussian, with only a few
parameters. The TGED therefore scales very favorably with $D$ and is
applicable to polyatomic molecules.

\section{Thawed Gaussian Ehrenfest dynamics\label{sec:TGED}}

To derive the TGED, let us assume that the nuclear state $\psi(t)$ in
Eq.~(\ref{eq:TDH_mol_ansatz}) is a Gaussian wavepacket
\begin{equation}
\psi(q,t)=\exp\left[  \frac{i}{\hbar}\left(  \frac{1}{2}x^{T}\cdot A_{t}\cdot
x+p_{t}^{T}\cdot x+\gamma_{t}\right)  \right]  , \label{eq:GWP}%
\end{equation}
where $x:=q-q_{t}$, in \textquotedblleft$A\gamma$\textquotedblright%
\ parametrization~\cite{Heller:1976,Patoz_Vanicek:2018,Lasser_Lubich:2020} or
by%
\begin{align}
\psi(q,t)  &  =\left(  \pi\hbar\right)  ^{-D/4}\left(  \det Q_{t}\right)
^{-1/2}\nonumber\\
&  \times\exp\left[  \frac{i}{\hbar}\left(  \frac{1}{2}x^{T}\cdot P_{t}\cdot
Q_{t}^{-1}\cdot x+p_{t}^{T}\cdot x+S_{t}\right)  \right]
\label{eq:GWP_Hagedorn}%
\end{align}
in \textquotedblleft$QPS$\textquotedblright%
\ parametrization.\cite{Heller:1976a,Hagedorn:1980,Hagedorn:1998,book_Lubich:2008,Ohsawa_Leok:2013,Lasser_Lubich:2020}
The real vector parameters $q_{t}$ and $p_{t}$ are the expectation values of
position and momentum. The complex symmetric $D\times D$ matrix $A_{t}$ with
positive-definite imaginary part controls the width of the Gaussian and
position-momentum correlation, while the real and imaginary parts of the
complex scalar $\gamma_{t}$ control, respectively, the phase and norm of the
wavepacket. In the $QPS$ parametrization, $Q_{t}$ and $P_{t}$ are complex
$D\times D$ matrices, which satisfy certain symplecticity
conditions,\cite{book_Lubich:2008} and $S_{t}$ is a real parameter
generalizing the classical action.

In the absence of coupling to electronic degrees of freedom, a Gaussian
nuclear wavepacket remains Gaussian if it is propagated with the standard
quadratic kinetic energy operator (\ref{eq:T_p}) and with a potential energy
operator that is at most quadratic in $q$, but that can depend both on time
and state.\cite{Vanicek:2023} Motivated by this observation, let us
approximate the potential energy matrix $\mathbf{V}(\hat{q})$ with an
effective state-dependent potential that is at most quadratic in $\hat{q}$.
While the nuclear mean-field potential $V_{n}(\hat{q};\mathbf{c}%
)=\mathbf{c}^{\dag}\mathbf{V}(\hat{q})\mathbf{c}$ in
Eq.~(\ref{eq:H_n_in_diab_basis}) already depends on the state of the system
via the electronic wavefunction $\mathbf{c}$, the effective potential
$\mathbf{V}_{\text{eff}}(\hat{q};\psi)$ may also depend on the nuclear state
$\psi$. Therefore, let us assume that%
\begin{equation}
\mathbf{V}(\hat{q})\approx\mathbf{V}_{\text{eff}}(\hat{q};\psi)=\mathbf{V}%
_{0}+\mathbf{V}_{1}{}^{T}\cdot\hat{x}+\hat{x}^{T}\cdot\mathbf{V}_{2}\cdot
\hat{x}/2, \label{eq:V_eff_mol}%
\end{equation}
where $\hat{x}=\hat{q}-q_{t}$ is a shifted position operator and
$\mathbf{V}_{0}$, $\mathbf{V}_{1}$, and $\mathbf{V}_{2}$ are possibly $\psi
$-dependent real $S\times S$ symmetric matrices of $D$-dimensional scalars,
vectors, and symmetric matrices, respectively. In this case, the mean-field
nuclear potential $V_{n}(\hat{q};\mathbf{c})$, appearing in the mean-field
nuclear TDSE (\ref{eq:TDH_TDSE_nu_separ}), will also be approximated as%
\begin{equation}
V_{n}(\hat{q};\mathbf{c})\approx V_{n,\text{eff}}(\hat{q};\psi,\mathbf{c}%
):=\mathbf{c}^{\dag}\mathbf{V}_{\text{eff}}(\hat{q};\psi)\mathbf{c}
\label{eq:V_mf_n_eff}%
\end{equation}
with an effective quadratic potential%
\begin{equation}
V_{n,\text{eff}}(\hat{q};\psi,\mathbf{c})=V_{n,0}+V_{n,1}{}^{T}\cdot\hat
{x}+\hat{x}^{T}\cdot V_{n,2}\cdot\hat{x}/2, \label{eq:V_eff_n}%
\end{equation}
where $V_{n,0}$, $V_{n,1}$, and $V_{n,2}$ are, respectively, the
$D$-dimensional scalar, vector, and symmetric matrix%
\begin{equation}
V_{n,j}:=\mathbf{c}^{\dag}\mathbf{V}_{j}\mathbf{c}. \label{eq:V_n_j}%
\end{equation}

Remarkably, the nuclear factor $\psi(t)$ of the Hartree product will remain
Gaussian because the effective mean-field nuclear potential (\ref{eq:V_eff_n})
is quadratic in $\hat{q}$. This will hold as long as the molecular potential
$\mathbf{V}(\hat{q})$ is approximated by a possibly state- and time-dependent,
but at most quadratic potential $\mathbf{V}_{\text{eff}}(\hat{q};\psi)$. We
thus obtain a family of \emph{thawed Gaussian Ehrenfest dynamics} methods,
which differ only by the choice of the quadratic effective potential
$\mathbf{V}_{\text{eff}}(\hat{q};\psi)$. As in the single-surface, purely
nuclear case,\cite{Vanicek:2023} one can consider the
variational,\cite{Fereidani_Vanicek:2023} local
harmonic,\cite{Kletnieks_Vanicek:2023a}
single-Hessian,\cite{Begusic_Vanicek:2019} global harmonic, local cubic
variational,\cite{Fereidani_Vanicek:2023a} single quartic
variational,\cite{Vanicek:2023} and other approximations; we shall do so in
the next section.

With the effective quadratic potential, the equations of motion for nuclei and
electrons become%
\begin{align}
i\hbar\dot{\psi}  &  =[T(\hat{p})+V_{n,\text{eff}}(\hat{q};\psi,\mathbf{c}%
)]\psi,\label{eq:TDSE_nu_GWP}\\
i\hbar\dot{\mathbf{c}}  &  =[\langle T(\hat{p})\rangle_{n}+\langle
\mathbf{V}_{\text{eff}}(\hat{q};\psi)\rangle_{n}]\mathbf{c}.
\label{eq:TDSE_el_GWP}%
\end{align}
Moreover, the nuclear expectation values of kinetic and potential energies,
appearing in the electronic TDSE (\ref{eq:TDSE_el_GWP}), can be evaluated
analytically as%
\begin{align}
\langle T(\hat{p})\rangle_{n}  &  =T(p_{t})+\operatorname{Tr}_{n}[m^{-1}%
\cdot\operatorname{Cov}(p)]/2,\label{eq:T_GWP}\\
\langle\mathbf{V}_{\text{eff}}(\hat{q};\psi)\rangle_{n}  &  =\mathbf{V}%
_{0}+\operatorname{Tr}_{n}[\mathbf{V}_{2}\cdot\operatorname{Cov}(q)]/2,
\label{eq:V_GWP}%
\end{align}
where $\operatorname{Tr}_{n}$ denotes a matrix trace over $D$ nuclear degrees
of freedom [$\operatorname{Tr}_{n}(A)=\sum_{j=1}^{D}A_{jj}$]
and~\cite{Vanicek:2023}
\begin{align}
\operatorname{Cov}(\hat{q})  &  =(\hbar/2)\left(  \operatorname{Im}%
A_{t}\right)  ^{-1}=(\hbar/2)Q_{t}\cdot Q_{t}^{\dag},\label{eq:Cov_q}\\
\operatorname{Cov}(\hat{p})  &  =(\hbar/2)A_{t}\cdot\left(  \operatorname{Im}%
A_{t}\right)  ^{-1}\cdot A_{t}^{\ast}=(\hbar/2)P_{t}\cdot P_{t}^{\dag}
\label{eq:Cov_p}%
\end{align}
are the position and momentum covariance matrices.

\begin{figure}
[htbp]%
\centering\includegraphics[width=\columnwidth]{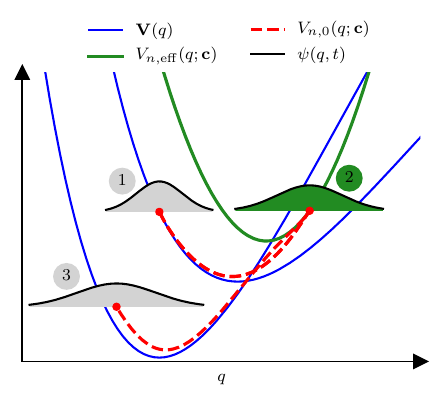}
\caption{{Example of thawed Gaussian Ehrenfest dynamics of the molecular wavepacket
$\Psi(q,t)=a(t)\psi(q,t)\varphi(t)$ moving in the molecular matrix potential $\mathbf{V}(q)$
[Eq.~(\ref{eq:V_matrix_els})], diagonal elements of which are shown as blue curves.  The nuclear
wavepacket $\psi(q,t)$ (black thin line) is
a Gaussian~(\ref{eq:GWP}) that exactly solves Eq.~(\ref{eq:TDSE_nu_GWP}) with an effective nuclear
potential $V_{n,\text{eff}}(\hat{q};\psi,\mathbf{c})$ [Eq.~(\ref{eq:V_eff_n}), indicated
by green thick line at instant 2], which is obtained from the effective molecular
potential~$\mathbf{V}_{\text{eff}}(\hat{q};\psi)$ by Eq.~(\ref{eq:V_mf_n_eff}), whose coefficients
are here given by the single-Hessian approximation [Eqs.~(\ref{eq:SH_TGED_a}) and (\ref{eq:SH_TGED_b}]
to the original, coupled
Morse system $\mathbf{V}(q)$. The electronic wavepacket $\varphi(t)$, expanded in the diabatic
basis [Eq.~(\ref{eq:phi_in_basis})], follows the coupled Eq.~(\ref{eq:TDSE_el_GWP}). The
red dashed line indicates the evolution of the zeroth-order coefficient $V_{n,0}$ of the nuclear
effective potential~(\ref{eq:V_eff_n})  from instant 1 to 3.}}\label{fig:TGED}
\end{figure}

As follows from the general thawed Gaussian wavepacket dynamics for a purely
nuclear wavepacket,\cite{Vanicek:2023} the nuclear TDSE (\ref{eq:TDSE_nu_GWP}%
)\ is equivalent to the system%
\begin{align}
\dot{q}_{t}  &  =m^{-1}\cdot p_{t},\label{eq:q_dot}\\
\dot{p}_{t}  &  =-V_{n,1},\label{eq:p_dot}\\
\dot{A}_{t}  &  =-A_{t}\cdot m^{-1}\cdot A_{t}-V_{n,2}\,,\label{eq:A_dot}\\
\dot{\gamma}_{t}  &  =T(p_{t})-V_{n,0}+(i\hbar/2)\operatorname*{Tr}\left(
m^{-1}\cdot A_{t}\right)  . \label{eq:gamma_dot}%
\end{align}
of ordinary differential equations for the Gaussian's parameters. In the $QPS$
parametrization, Eqs.~(\ref{eq:A_dot}) and (\ref{eq:gamma_dot}) for $A_{t}$
and $\gamma_{t}$ are replaced with%
\begin{align}
\dot{Q}_{t}  &  =m^{-1}\cdot P_{t},\label{eq:Q_dot}\\
\dot{P}_{t}  &  =-V_{n,2}\cdot Q_{t},\label{eq:P_dot}\\
\dot{S}_{t}  &  =T(p_{t})-V_{n,0}. \label{eq:S_dot}%
\end{align}
The only difference from the equations for a purely nuclear wavepacket
\cite{Vanicek:2023} is that Eqs.~(\ref{eq:q_dot})--(\ref{eq:S_dot}) are
coupled to the electronic propagation~(\ref{eq:TDSE_el_GWP}) via the
coefficients $V_{n,j}$ defined in Eq.~(\ref{eq:V_n_j}) from the effective
potential~(\ref{eq:V_eff_mol}). Figure~\ref{fig:TGED} visualizes the TGED by
displaying a trajectory of the nuclear wavepacket $\psi(q,t)$ as well as the
original and effective potentials.

From the construction of the TGED, it is clear that the method conserves the
norm of both nuclear and electronic states, i.e., $\left\Vert \psi
_{t}\right\Vert =\left\Vert \mathbf{c}_{t}\right\Vert =1$. This can also be
seen explicitly from the Hermitian property of both electronic and nuclear
mean-field Hamiltonians. Because the prefactor $a(t)$ is a complex unit, the
norm of the molecular wavefunction is also conserved: $\left\Vert
\Psi(t)\right\Vert =1$. By construction, the TGED is reversible, as are the
thawed Gaussian wavepacket and Ehrenfest dynamics individually. When the
effective quadratic potential is obtained by the variational principle, also
the exact energy is conserved; the resulting \textquotedblleft variational
TGED\textquotedblright\ is discussed in Sec.~\ref{sec:V-TGED}.

\section{Family of thawed Gaussian Ehrenfest dynamics
methods\label{sec:TGED_family}}

By choosing different effective locally quadratic
potentials~(\ref{eq:V_eff_mol}), one obtains different thawed Gaussian
Ehrenfest dynamics methods. As in the single-surface case,\cite{Vanicek:2023}
there is an infinite family of such methods. In the nonadiabatic setting, the
effective potential ~(\ref{eq:V_eff_mol}) can be thought of as a local\ and
state-dependent variant of the popular second-order vibronic coupling
model.\cite{Koppel_Cederbaum:1984,book_Domcke_Koppel:2004}

\subsection{Variational thawed Gaussian Ehrenfest dynamics\label{sec:V-TGED}}

The most accurate TGED method is obtained by applying the variational
principle. Remarkably, for a fully variational treatment of the Hartree
product of an electronic and Gaussian nuclear\ wavepackets, one does not need
to do any extra work. The electronic TDSE (\ref{eq:TDSE_el_GWP}) is already
fully variational, while the mean-field nuclear TDSE (\ref{eq:TDSE_nu_GWP})
can be thought of as a standard\ nuclear TDSE driven by a time-dependent
external\ potential%
\begin{equation}
V_{n}(\hat{q},t):=V_{n}(\hat{q};\mathbf{c}_{t})\equiv\mathbf{c}_{t}^{\dag
}\mathbf{V}(\hat{q})\mathbf{c}_{t}\mathbf{.}%
\end{equation}
Applying the TDVP\ to the mean-field nuclear TDSE (\ref{eq:TDH_TDSE_nu_separ})
with a Gaussian ansatz (\ref{eq:GWP}) or (\ref{eq:GWP_Hagedorn}) yields a
quadratic effective nuclear potential%
\begin{equation}
V_{n,\text{var}}(\hat{q})=V_{n,0}+V_{n,1}{}^{T}\cdot\hat{x}+\hat{x}^{T}\cdot
V_{n,2}\cdot\hat{x}/2,
\end{equation}
with scalar, vector, and matrix coefficients%
\begin{align}
V_{n,0} &  =\langle V_{n}(\hat{q},t)\rangle_{n}-\operatorname{Tr}_{n}[\langle
V_{n}^{\prime\prime}(\hat{q},t)\rangle_{n}\cdot\operatorname{Cov}(q)]/2,\text{
}\\
V_{n,1} &  =\langle V_{n}^{\prime}(\hat{q},t)\rangle_{n},\\
V_{n,2} &  =\langle V_{n}^{\prime\prime}(\hat{q},t)\rangle_{n},
\end{align}
as shown for the variational TGWD in the one-surface
case.\cite{Coalson_Karplus:1990,Vanicek:2023} To find the coefficients
$\mathbf{V}_{j}$ of the effective molecular potential $\mathbf{V}_{\text{var}%
}(\hat{q})$ in Eq.~(\ref{eq:V_eff_mol}), we note that%
\begin{equation}
\langle V_{n}^{(j)}(\hat{q},t)\rangle_{n}=\mathbf{c}_{t}^{\dag}\langle
\mathbf{V}^{(j)}(\hat{q})\rangle_{n}\mathbf{c}_{t}=\langle\mathbf{V}%
^{(j)}(\hat{q})\rangle.
\end{equation}
Hence $V_{n,j}=\mathbf{c}_{t}^{\dag}\mathbf{V}_{j}\mathbf{c}_{t}$, as in
Eq.~(\ref{eq:V_n_j}), where%
\begin{align}
\mathbf{V}_{0} &  =\langle\mathbf{V}(\hat{q})\rangle_{n}-\operatorname{Tr}%
_{n}[\langle\mathbf{V}^{\prime\prime}(\hat{q})\rangle_{n}\cdot
\operatorname{Cov}(q)]/2,\text{ }\label{eq:V_mol_VGA_0}\\
\mathbf{V}_{1} &  =\langle\mathbf{V}^{\prime}(\hat{q})\rangle_{n}%
,\label{eq:V_mol_VGA_`}\\
\mathbf{V}_{2} &  =\langle\mathbf{V}^{\prime\prime}(\hat{q})\rangle
_{n}.\label{eq:V_mol_VGA_2}%
\end{align}
We have thus obtained the \emph{variational thawed Gaussian Ehrenfest
dynamics}, which is a generalization of Heller's and Coalson and Karplus's
variational thawed Gaussian wavepacket
dynamics\cite{Heller:1976,Coalson_Karplus:1990} to nonadiabatic systems.

\subsection{Local harmonic thawed Gaussian Ehrenfest
dynamics\label{sec:LH-TGED}}

Evaluating expectation values $\langle\mathbf{V}^{(j)}(\hat{q})\rangle_{n}$,
needed in the variational TGED may be difficult in practical calculations with
non-polynomial potential energy surfaces. In the limit of small $\hbar$, it
pays off to approximate the potential energy by its local harmonic
(LH)\ approximation, a quadratic\ expansion about the center $q_{t}$ of the
wavepacket, as was done by Heller in the single-surface
setting.\cite{Heller:1975} Here, one simply replaces matrix coefficients
$\mathbf{V}_{j}$ in Eq.~(\ref{eq:V_eff_mol}) with the diabatic potential
energy matrix, its gradient, and Hessian at the wavepacket's center:%
\begin{equation}
\mathbf{V}_{j}=\mathbf{V}^{(j)}(q_{t})\text{ \ \ for }j=0,1,2
\label{eq:V_LHA_j}%
\end{equation}
The mean-field potential energy operator (\ref{eq:V_GWP}) needed in the
electronic TDSE (\ref{eq:TDSE_el_GWP}) then becomes%
\begin{equation}
\langle\mathbf{V}_{\text{LH}}(\hat{q};\psi)\rangle_{n}=\mathbf{V}\left(
q_{t}\right)  +\operatorname{Tr}_{n}[\mathbf{V}^{\prime\prime}\left(
q_{t}\right)  \cdot\operatorname{Cov}(q)]/2.
\end{equation}
Although Eq.~(\ref{eq:V_LHA_j}) together with the general TGED
Eqs.~(\ref{eq:TDSE_el_GWP}) and (\ref{eq:q_dot})-(\ref{eq:S_dot}) describe the
local harmonic version of TGED completely, let us write the new equations of
motion for the electronic wavefunction and nuclear position and momentum
explicitly:%
\begin{align}
i\hbar\dot{\mathbf{c}}_{t}  &  =[\langle T(\hat{p})\rangle_{n}+\langle
\mathbf{V}_{\text{LH}}(\hat{q};\psi)\rangle_{n}]\mathbf{c}_{t},\\
\dot{q}_{t}  &  =m^{-1}\cdot p_{t},\\
\dot{p}_{t}  &  =-V_{n}^{\prime}(q_{t}).
\end{align}
The width and phase of the nuclear wavepacket evolve according to the
equations%
\begin{align}
\dot{A}_{t}  &  =-A_{t}\cdot m^{-1}\cdot A_{t}-V_{n}^{\prime\prime}(q_{t}),\\
\dot{\gamma}_{t}  &  =T(p_{t})-V_{n}(q_{t})+(i\hbar/2)\operatorname*{Tr}%
\left(  m^{-1}\cdot A_{t}\right)
\end{align}
in the $A\gamma$ parametrization and according to equations%
\begin{align}
\dot{Q}_{t}  &  =m^{-1}\cdot P_{t},\\
\dot{P}_{t}  &  =-V_{n}^{\prime\prime}(q_{t})\cdot Q_{t},\\
\dot{S}_{t}  &  =T(p_{t})-V_{n}(q_{t})
\end{align}
in the $QPS$ parametrization. This system of differential equations expresses
the \emph{local harmonic thawed Gaussian Ehrenfest dynamics}, which is a
generalization of Heller's original thawed Gaussian
approximation\cite{Heller:1975} to the setting with multiple electronic states.

\subsection{Global harmonic thawed Gaussian Ehrenfest
dynamics\label{sec:H-TGED}}

A cruder, yet more efficient, \emph{harmonic thawed Gaussian Ehrenfest
dynamics} is obtained by replacing the molecular potential energy operator
$\mathbf{V}(\hat{q})$ with its global harmonic approximation%
\begin{equation}
\mathbf{V}_{\text{harm}}(\hat{q})=\mathbf{V}\left(  q_{r}\right)
+\mathbf{V}^{\prime}\left(  q_{r}\right)  ^{T}\cdot\hat{x}_{r}+\hat{x}_{r}%
^{T}\cdot\mathbf{V}^{\prime\prime}\left(  q_{r}\right)  \cdot\hat{x}_{r}/2,
\label{eq:V_mol_HA}%
\end{equation}
where $\hat{x}_{r}:=\hat{q}-q_{r}$ is the displacement from a a fixed
reference position $q_{r}$.\ This is equivalent to setting the coefficients of
the effective molecular potential (\ref{eq:V_eff_mol}) to%
\begin{align}
\mathbf{V}_{0}  &  =\mathbf{V}_{\text{harm}}(q_{t}),\\
\mathbf{V}_{1}  &  =\mathbf{V}_{\text{harm}}^{\prime}(q_{t})=\mathbf{V}%
^{\prime}(q_{r})+\mathbf{V}^{\prime\prime}\left(  q_{r}\right)  \cdot
(q_{t}-q_{r}),\\
\mathbf{V}_{2}  &  =\mathbf{V}_{\text{harm}}^{\prime\prime}(q_{t}%
)=\mathbf{V}^{\prime\prime}(q_{r}).
\end{align}
Harmonic approximation (\ref{eq:V_mol_HA}) to the molecular potential is an
example of the quadratic vibronic coupling
model,\cite{Koppel_Cederbaum:1984,book_Domcke_Koppel:2004} widely used for
nonadiabatic simulations.

\subsection{Single-Hessian thawed Gaussian Ehrenfest
dynamics\label{sec:SH-TGED}}

To include some anharmonicity beyond the harmonic TGED but avoid the costly
evaluation of the Hessians needed in the local harmonic TGED, one can
generalize the single-Hessian thawed Gaussian wavepacket
dynamics\cite{Begusic_Vanicek:2019,Begusic_Vanicek:2022,Barbiero_Vanicek:2026}
and obtain the \emph{single-Hessian thawed Gaussian Ehrenfest dynamics}, where
the coefficients of the effective potential (\ref{eq:V_eff_mol}) are set to%
\begin{align}
\mathbf{V}_{j} &  =\mathbf{V}^{(j)}(q_{t})\text{ \ \ for }j=0\text{ and
}1,\label{eq:SH_TGED_a}\\
\mathbf{V}_{2} &  =\mathbf{V}^{\prime\prime}(q_{r}).\label{eq:SH_TGED_b}%
\end{align}
The effective potential has the exact value and gradient but its Hessian
(curvature) is kept constant. Besides its efficiency, in the one-surface
setting the single-Hessian TGWD was shown to conserve both the symplectic
structure and effective
energy.\cite{Begusic_Vanicek:2019,Barbiero_Vanicek:2026}

\subsection{Local cubic variational TGED\label{sec:LCV-TGED}}

To include effects beyond the local harmonic approximation, it is useful to
apply the variational TGED to a local cubic approximation for the potential
and obtain the \emph{local cubic variational thawed Gaussian Ehrenfest
dynamics}, for which the coefficients of $\mathbf{V}_{\text{eff}}$ are
\begin{align}
\mathbf{V}_{j} &  =\mathbf{V}^{(j)}(q_{t})\text{ \ \ for }j=0\text{ and }2,\\
\mathbf{V}_{1,k} &  =\mathbf{V}^{\prime}(q_{t})_{k}+\frac{1}{2}\sum
_{l,m=1}^{D}\mathbf{V}^{\prime\prime\prime}(q_{t})_{klm}\operatorname{Cov}%
(q)_{lm}.
\end{align}
In the single-surface setting, this approximation (called \textquotedblleft
extended semiclassical,\textquotedblright\ \textquotedblleft symplectic
semiclassical,\textquotedblright\ or \textquotedblleft local cubic
variational\textquotedblright%
\ TGWD),\cite{Pattanayak_Schieve:1994,Ohsawa_Leok:2013,Fereidani_Vanicek:2023a}
has been shown to conserve the symplectic structure, effective energy, and can
qualitatively capture tunneling without requiring the expectation values of
the potential energy derivatives needed in the variational TGWD.

\subsection{Single quartic variational TGED\label{sec:SQV-TGED}}

The limitation of the local cubic variational TGED is that the local cubic
potential is necessarily unbounded from below. To avoid the resulting
numerical issues but keep approximately the same computational cost, the
single-quartic variational TGWD was proposed in the single-surface
setting.\cite{Vanicek:2023} In this method, the variational principle is
applied to the single-quartic approximation of the potential, which has the
exact local derivative up to the third order but keeps a single constant
positive definite fourth derivative tensor. Remarkably, like the
single-Hessian variant, this method is symplectic and conserves the effective
energy, neither of which is true for the local harmonic or the much more
expensive local quartic approximations. Generalizing the results to the
nonadiabatic setting, the coefficients of the effective potential of the
\emph{single-quartic variational TGED }are
\begin{align}
\mathbf{V}_{2,}{}_{ij} &  =\mathbf{V}^{\prime\prime}(q_{t})_{ij}+\sum
_{k,l=1}^{D}\mathbf{V}^{(4)}(q_{r})_{ijkl}\Sigma_{kl}/2,\label{eq:V2_SQV-TGED}%
\\
\mathbf{V}_{1,i} &  =\mathbf{V}^{\prime}(q_{t})_{i}+\sum_{j,k=1}^{D}%
\mathbf{V}^{\prime\prime\prime}(q_{t})_{ijk}\Sigma_{jk}%
/2,\label{eq:V1_SQV-TGED}\\
\mathbf{V}_{0} &  =\mathbf{V}(q_{t})-\sum_{i,j,k,l=1}^{D}\mathbf{V}%
^{(4)}(q_{r})_{ijkl}\Sigma_{ij}\Sigma_{kl}/8,\label{eq:V0_SQV-TGED}%
\end{align}
where $\Sigma\equiv\operatorname{Cov}(q)$ is the shorthand notation for the
position covariance matrix (\ref{eq:Cov_q}). Because only a single fourth
derivative is needed, if this derivative is evaluated by finite differences,
the increase in cost over the local cubic variational TGED would be negligible
in typical simulations, where the number of time steps is much larger than the
number of degrees of freedom. Because the single-quartic potential can always
be bounded from below, the single-quartic variational TGWD appears to be the
method of choice if one wants to improve accuracy beyond the local harmonic
TGED but avoid the cost of the fully variational TGED.

\section{Special limits of TGED\label{sec:limits_of_TGED}}

\begin{figure}
[htbp]%
\centering\includegraphics[width=\columnwidth]{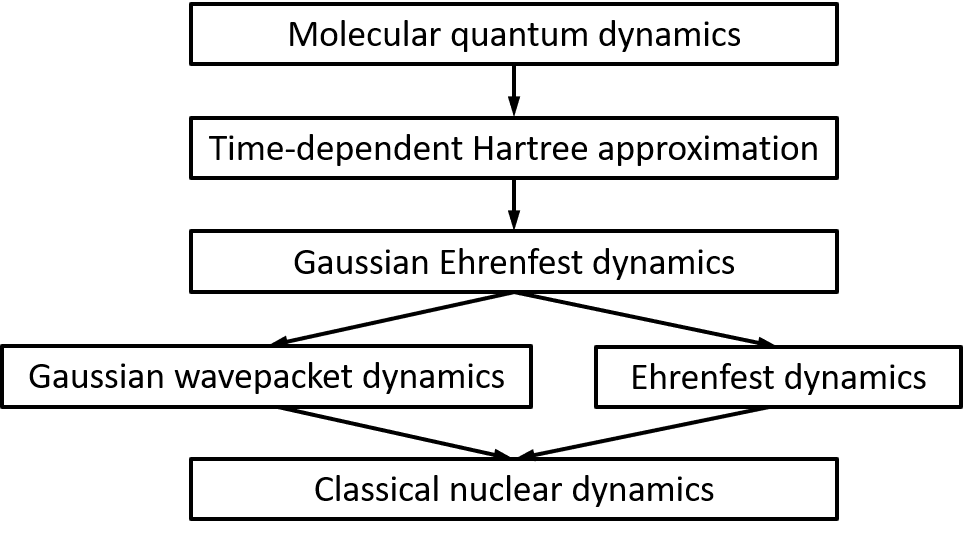}
\caption{{Relation of thawed Gaussian Ehrenfest dynamics to exact
molecular quantum dynamics and to other approximations.}}\label{fig:diagram}
\end{figure}

Let us explore various limits of the thawed Gaussian Ehrenfest dynamics (see
Fig.~\ref{fig:diagram}). Note that in Secs.~\ref{sec:limit_uncoupled},
\ref{sec:limit_TGWD}, and \ref{sec:limit_classical_dynamics}, $c_{s}(t)$
denotes the component of the electronic vector $\mathbf{c}(t)$ in the $s$th
electronic state; the time $t$ is therefore not in the subscript (as in
$\mathbf{c}_{t}$), but in the argument. Comparison of the TGED with three of
its limits (Ehrenfest dynamics, TGWD, and classical dynamics) in a simple
system, where TGED is exact, was shown in Fig.~\ref{fig:TGED_comparison}.

\subsection{TGED with constant electronic
populations\label{sec:limit_uncoupled}}

If the electronic states are not coupled, $\mathbf{V}(q)$ is a diagonal matrix
[$V_{rs}(q)=0$ for $r\neq s$], the electronic phase on each surface evolves
independently, the population of each electronic state remains constant, and
the mean-field potential for nuclear motion is determined by the initial
populations. Equations (\ref{eq:TDH_TDSE_nu_separ}%
)-(\ref{eq:TDH_TDSE_el_separ}) reduce to%
\begin{align}
i\hbar\dot{\psi}  &  =[T(\hat{p})+\sum_{s=1}^{S}V_{ss}(\hat{q})|c_{s}%
(t=0)|^{2}]\psi,\label{eq:TDH_TDSE_nu_uncpl}\\
i\hbar\dot{c}_{s}  &  =[\langle T(\hat{p})\rangle_{n}+\langle V_{ss}(\hat
{q})\rangle_{n}]c_{s},~~~s=1,\ldots,S \label{eq:TDH_TDSE_el_uncpl}%
\end{align}
The nuclear wavepacket still feels the mean-field potential energy, but the
weights of all surfaces remain unchanged. [To derive
Eqs.~(\ref{eq:TDH_TDSE_nu_uncpl}) and (\ref{eq:TDH_TDSE_el_uncpl}), we did not
have to assume the nuclear wavepacket to be Gaussian.]

\subsection{Thawed Gaussian wavepacket dynamics\label{sec:limit_TGWD}}

If the electronic states are uncoupled, and, in addition, the electrons are
initially in a single state $s$ [$c_{r}(0)=\delta_{rs}$], the preceding
equations further reduce to%
\begin{align}
i\hbar\dot{\psi}  &  =[T(\hat{p})+V_{ss}(\hat{q})]\psi,\\
i\hbar\dot{c}_{s}  &  =[\langle T(\hat{p})\rangle_{n}+\langle V_{ss}(\hat
{q})\rangle_{n}]c_{s}=Ec_{s}.
\end{align}
Therefore $c_{s}(t)=\exp(-iEt/\hbar)$, and the cancellation between the
electronic coefficient $c_{s}(t)$ and the prefactor $a(t)=\exp(iEt/\hbar)$
reduces the molecular state to%
\begin{equation}
\Psi(t)=a(t)\psi(t)c_{s}(t)|s\rangle=\psi(t)|s\rangle.
\end{equation}
In other words, dynamics reduces to Born-Oppenheimer nuclear dynamics on a
single surface $s$. Moreover, because the nuclear wavepacket $\psi(t)$ in TGED
is Gaussian, one obtains the thawed Gaussian wavepacket dynamics. Depending on
the choice of the molecular effective potential used in TGED, one obtains the
corresponding member of the TGWD family, such as the variational, local
harmonic, single-Hessian, global harmonic, local cubic variational, or the
single quartic variational TGWD.\cite{Vanicek:2023}

\subsection{Ehrenfest dynamics\label{sec:limit_Ehrenfest}}

The classical limit for nuclei is achieved by letting the effective Planck
constant approach zero, $\hbar\rightarrow0$. In this limit, the position and
momentum covariances [Eqs.~(\ref{eq:Cov_q}) and (\ref{eq:Cov_p})], which are
proportional to $\hbar$, vanish, and as a result, the contributions to
$\mathbf{V}_{\text{var},j}$, $\langle T(\hat{p})\rangle_{n}$, and
$\langle\mathbf{V}^{(j)}(\hat{q})\mathbf{\rangle}_{n}$ from the finite width
of the wavepacket become negligible in Eqs.~(\ref{eq:V_mol_VGA_0}%
)--(\ref{eq:V_mol_VGA_2}) and in Eqs.~(\ref{eq:T_GWP}) and (\ref{eq:V_GWP}).
Because the wavepacket becomes increasingly localized, the classical limit can
be accomplished formally by using a \emph{local linear (LL) approximation} for
the potential and kinetic energies:%
\begin{align}
\mathbf{V}(\hat{q})  &  \approx\mathbf{V}_{\text{LL}}(\hat{q})=\mathbf{V}%
\left(  q_{t}\right)  +\mathbf{V}^{\prime}\left(  q_{t}\right)  ^{T}\cdot
(\hat{q}-q_{t}),\label{eq:V_LLA}\\
T(\hat{p})  &  \approx T_{\text{LL}}(\hat{p})=T\left(  p_{t}\right)
+T^{\prime}\left(  p_{t}\right)  ^{T}\cdot(\hat{p}-p_{t}).
\end{align}
With this approximation, the effective potential coefficients of the
variational TGED\ and the expectation values of kinetic and potential energies
become%
\begin{align}
\mathbf{V}_{\text{var},2}  &  \overset{\hbar\rightarrow0}{\rightarrow
}\mathbf{V}_{\text{LL},2}=0,\\
\mathbf{V}_{\text{var,}1}  &  \overset{\hbar\rightarrow0}{\rightarrow
}\mathbf{V}_{\text{LL},1}=\mathbf{V}^{\prime}(q_{t}),\\
\mathbf{V}_{\text{var,}0}  &  \overset{\hbar\rightarrow0}{\rightarrow
}\mathbf{V}_{\text{LL,}0}=\mathbf{V}(q_{t}),\\
\langle T(\hat{p})\rangle_{n}  &  \overset{\hbar\rightarrow0}{\rightarrow
}\langle T_{\text{LL}}(\hat{p})\rangle_{n}=T(p_{t}),\label{eq:T_GWP_small_h}\\
\langle\mathbf{V}_{\text{var}}(\hat{q})\rangle_{n}  &  \overset{\hbar
\rightarrow0}{\rightarrow}\langle\mathbf{V}_{\text{LL}}(\hat{q})\rangle
_{n}=\mathbf{V}(q_{t}). \label{eq:V_GWP_small_h}%
\end{align}

The electronic TDSE (\ref{eq:TDSE_el_GWP}) and Eqs.~(\ref{eq:q_dot}) and
(\ref{eq:p_dot}) for nuclear positions and momenta reduce to the equations of
standard mixed quantum-classical \emph{Ehrenfest dynamics}:%
\begin{align}
i\hbar\dot{\mathbf{c}}_{t}  &  =[T(p_{t})+\mathbf{V}(q_{t})]\mathbf{c}%
_{t},\label{eq:c_dot_Ehr_diab}\\
\dot{q}_{t}  &  =m^{-1}\cdot p_{t},\label{eq:q_dot_Ehr_diab}\\
\dot{p}_{t}  &  =-V_{n}^{\prime}(q_{t}). \label{eq:p_dot_Ehr_diab}%
\end{align}
As for the evolution of the other parameters,%
\begin{align}
\dot{A}_{t}  &  =\dot{Q}_{t}=\dot{P}_{t}=0,\\
\dot{\gamma}_{t}  &  =\dot{S}_{t}=T(p_{t})-V_{n}(q_{t}).
\end{align}
The Gaussian wavepacket becomes \textquotedblleft frozen\textquotedblright%
\ because its width matrix remains constant ($A_{t}=A_{0}$ or $Q_{t}=Q_{0}$
and $P_{t}=P_{0}$). Moreover, the increment of $\gamma_{t}$ along the
trajectory is equal to the increment of Ehrenfest-averaged classical action
$S_{t}$. Although the nuclear action $S_{t}$ is usually ignored in Ehrenfest
dynamics, propagating $S_{t}$ is important if one is interested in the
time-dependent overall phase of the wavepacket needed, e.g., in the
calculations of spectra. The conserved total energy is%
\begin{equation}
E=\mathbf{c}_{t}^{\dag}[T(p_{t})+\mathbf{V}(q_{t})]\mathbf{c}_{t}%
=T(p_{t})+V_{n}(q_{t}), \label{eq:E_Ehrenfest}%
\end{equation}
which also reflects that the wavepacket width is ignored in the classical limit.

\subsection{Classical nuclear dynamics\label{sec:limit_classical_dynamics}}

By taking the classical limit of TGWD or the single-surface limit (on surface
$s$) of Ehrenfest dynamics, one obtains the single-surface classical nuclear
dynamics%
\begin{align}
c_{s}(t)  &  =\exp(-iEt/\hbar),\\
\dot{q}_{t}  &  =m^{-1}\cdot p_{t},\\
\dot{p}_{t}  &  =-V_{ss}^{\prime}(q_{t}),\\
\dot{A}_{t}  &  =\dot{Q}_{t}=\dot{P}_{t}=0,\\
\dot{\gamma}_{t}  &  =\dot{S}_{t}=T(p_{t})-V_{ss}(q_{t}),
\end{align}
where $E=T(p_{t})+V_{ss}(q_{t})$ is the conserved classical energy on the
$s$th surface.

\subsection{Independent electronic dynamics and free-particle nuclear thawed
Gaussian wavepacket dynamics}

Now let us consider the limit in which the potential energy operator does not
depend on nuclear coordinates:%
\begin{equation}
\mathbf{V}(\hat{q})\approx\mathbf{V}=\operatorname{const}.
\end{equation}
The electronic TDSE (\ref{eq:TDSE_el_GWP}) can be solved exactly, giving%
\begin{equation}
\mathbf{c}_{t}=\exp\{-it[\langle T(\hat{p})\rangle_{n,t=0}\mathbf{1}%
+\mathbf{V}]/\hbar\}\mathbf{c}_{0}.
\end{equation}
The mean-field nuclear potential (\ref{eq:V_mf_n_eff}) is%
\begin{equation}
V_{n}=\mathbf{c}_{t}^{\dag}\mathbf{Vc}_{t}=\mathbf{c}_{0}^{\dag}%
\mathbf{Vc}_{0}=V_{n,t=0}%
\end{equation}
since the exponential in $\mathbf{c}_{t}$ commutes with $\mathbf{V}$.
Equations for Gaussian parameters have exact solutions\cite{Vanicek:2023}%
\begin{align}
q_{t} &  =q_{0}+tm^{-1}\cdot p_{0},\\
p_{t} &  =p_{0}=\operatorname{const},\\
A_{t} &  =A_{0}\cdot\left(  \operatorname{Id}_{D}+tm^{-1}\cdot A_{0}\right)
^{-1},\\
\gamma_{t} &  =\gamma_{0}+t[T(p_{0})-V_{n,t=0}]\nonumber\\
&  ~~~+(i\hbar/2)\ln\det\left(  \operatorname{Id}_{D}+tm^{-1}\cdot
A_{0}\right)  ,\\
Q_{t} &  =Q_{0}+tm^{-1}\cdot P_{0},\\
P_{t} &  =P_{0}=\operatorname{const},\\
S_{t} &  =S_{0}+t[T(p_{0})-V_{n,t=0}].
\end{align}
Equations for the position, momentum, and width matrix ($A_{t}$ or $Q_{t}$ and
$P_{t}$) are the same as for a Gaussian free particle (see
Sec.~\ref{sec:TGED_T_prop} on kinetic propagation), but the equation for
$\gamma_{t}$ (or $S_{t}$) contains a potential term $V_{n}$.

\section{Geometric integrators\label{sec:integrators}}

Not all numerical integration schemes preserve the geometric properties of the
exact or approximate solution of the molecular Schr\"{o}dinger
equation.\cite{book_Hairer_Wanner:2006} There is, however, a general strategy
to preserve energy conservation approximately and all other properties
exactly. If the Hamiltonian $\mathcal{H}=\mathcal{T}+\mathcal{V}$ is separable
into two terms, $\mathcal{T}$ and $\mathcal{V}$, the propagation with each of
which can be performed exactly, one can obtain structure-preserving
integrators of arbitrary even order of accuracy by symmetrically composing the
Strang splitting of the evolution operator.\cite{book_Hairer_Wanner:2006} Such
integrators were obtained for the exact quantum solution of the nonadiabatic
Schr\"{o}dinger equation in
Refs.~\onlinecite{Choi_Vanicek:2019,Roulet_Vanicek:2019}, for the
representation-free Ehrenfest dynamics in
Ref.~\onlinecite{Choi_Vanicek:2021a}, and for the generalized TGWD in
Ref.~\onlinecite{Vanicek:2023}. Using the same procedure, one can obtain
geometric integrators of arbitrary even order of accuracy for the TGED. In
particular, we only need to find the exact solutions of the ordinary
differential equations for the kinetic and potential propagations of the
Gaussian's parameters $q_{t}$, $p_{t},A_{t}$, $\gamma_{t}$, $Q_{t}$, $P_{t}$,
and $S_{t}$ and for the electronic wavefunction $\mathbf{c}_{t}$. We do so next.

\subsection{Kinetic propagation\label{sec:TGED_T_prop}}

During the kinetic propagation step, the effective Hamiltonian $\mathbf{\hat
{H}}_{\text{eff}}=T(\hat{p})\mathbf{1}$ contains only the kinetic energy. As a
result, the electronic TDSE becomes%
\begin{equation}
i\hbar\mathbf{\dot{c}}_{t}=T(\psi_{t})\mathbf{c}_{t},
\end{equation}
where the nuclear kinetic energy $T(\psi_{t}):=\langle T(\hat{p})\rangle_{n}$
is given by Eq.~(\ref{eq:T_GWP}), while the equations of motion for the
nuclear wavepacket reduce to%
\begin{align}
\dot{q}_{t}  &  =m^{-1}\cdot p_{t},\\
\dot{p}_{t}  &  =0,\\
\dot{A}_{t}  &  =-A_{t}\cdot m^{-1}\cdot A_{t}\,,\\
\dot{\gamma}_{t}  &  =T(p_{t})+(i\hbar/2)\operatorname{Tr}_{n}(m^{-1}\cdot
A_{t}),\\
\dot{Q}_{t}  &  =m^{-1}\cdot P_{t},\\
\dot{P}_{t}  &  =0,\\
\dot{S}_{t}  &  =T(p_{t}).
\end{align}

To solve the electronic TDSE analytically, we note that not only the momentum
covariance (\ref{eq:Cov_p}) but also the kinetic energy (\ref{eq:T_GWP})
remains constant since neither $p_{t}$ nor $P_{t}$ evolves during the kinetic
propagation. Hence the propagation of the electronic wavefunction is a simple
multiplication with a scalar exponential:%
\begin{equation}
\mathbf{c}_{t}=\exp[-itT(\psi_{0})/\hbar]\mathbf{c}_{0}.
\end{equation}
Analytical solution for the Gaussian parameters is the same as in the
single-surface TGWD:\cite{Vanicek:2023}%
\begin{align}
q_{t}  &  =q_{0}+tm^{-1}\cdot p_{0},\label{eq:q_t_T_prop}\\
p_{t}  &  =p_{0},\label{eq:p_t_T_prop}\\
A_{t}  &  =\left(  A_{0}^{-1}+tm^{-1}\right)  ^{-1}=A_{0}\cdot\left(
\operatorname{Id}_{D}+tm^{-1}\cdot A_{0}\right)  ^{-1}\nonumber\\
&  =\left(  \operatorname{Id}_{D}+tA_{0}\cdot m^{-1}\right)  ^{-1}\cdot
A_{0},\label{eq:A_t_T_prop}\\
\gamma_{t}  &  =\gamma_{0}+tT(p_{0})+\frac{i\hbar}{2}\ln\det\left(
\operatorname{Id}_{D}+tm^{-1}\cdot A_{0}\right)  ,\label{eq:gamma_t_T_prop}\\
Q_{t}  &  =Q_{0}+tm^{-1}\cdot P_{0},\\
P_{t}  &  =P_{0},\\
S_{t}  &  =S_{0}+tT(p_{0}).
\end{align}

\subsection{Potential propagation\label{sec:TGED_V_prop}}

During the potential propagation step, the effective Hamiltonian
$\mathbf{\hat{H}}_{\text{eff}}=\mathbf{V}_{\text{eff}}(\hat{q};\psi)$ contains
only the potential energy. Therefore, the electronic TDSE becomes%
\begin{equation}
i\hbar\mathbf{\dot{c}}_{t}=\mathbf{V}_{\text{eff}}(\psi_{t})\mathbf{c}_{t},
\end{equation}
where the electronic potential energy matrix $\mathbf{V}_{\text{eff}}%
(\psi):=\langle\mathbf{V}_{\text{eff}}(\hat{q};\psi)\rangle_{n}$ is given by
Eq.~(\ref{eq:V_GWP}), while the nuclear Gaussian wavepacket evolves according
to the equations%
\begin{align}
\dot{q}_{t}  &  =0\\
\dot{p}_{t}  &  =-V_{n,1},\\
\dot{A}_{t}  &  =-V_{n,2}\,,\\
\dot{\gamma}_{t}  &  =-V_{n,0},\\
\dot{Q}_{t}  &  =0,\\
\dot{P}_{t}  &  =-V_{n,2}\cdot Q_{t},\\
\dot{S}_{t}  &  =-V_{n,0}.
\end{align}

The electronic equation is easy to solve if $\mathbf{V}_{j}$ does not change
during the potential propagation. Since $\mathbf{V}_{\text{eff}}$ is assumed
to depend only on the nuclear and not on the electronic state, we only need to
worry about the dependence of $\mathbf{V}_{j}$ on the Gaussian wavepacket's
parameters. As we now show, $\mathbf{V}_{j}$ will not depend on time as long
as $\mathbf{V}_{j}$ depends only on parameters $q_{t}$ and $Q_{t}$ (or $q_{t}$
and $\operatorname{Im}A_{t}$). Note that this assumption holds for all
discussed methods: e.g., $\mathbf{V}_{\text{LH,}j}$ depends only on $q_{t}$
and $\mathbf{V}_{\text{var,}j}$ depends only on $q_{t}$ and $Q_{t}$ or
$\operatorname{Im}A_{t}$ (which appear in the coordinate-space density needed
in evaluating expectation values). As neither $q_{t}$ nor $Q_{t}$ changes
during the potential propagation, $\mathbf{V}_{j}$ and $\mathbf{V}%
_{\text{eff}}$ also remain constant. The electronic wavefunction is thus
obtained by multiplying the initial state with a matrix exponential:%
\begin{equation}
\mathbf{c}_{t}=\exp[-it\mathbf{V}_{\text{eff}}(\psi_{0})/\hbar]\mathbf{c}_{0}.
\label{eq:c_t_V_prop}%
\end{equation}

Although $\mathbf{V}_{j}$ remains constant during the potential propagation,
the nuclear effective potential energy coefficients $V_{n,j}$ from
Eq.~(\ref{eq:V_n_j}) change due to their dependence on the electronic
wavefunction $\mathbf{c}_{t}$, which evolves according to
Eq.~(\ref{eq:c_t_V_prop}). The evolved Gaussian's parameters can be written
as
\begin{align}
q_{t}  &  =q_{0},\\
p_{t}  &  =p_{0}-t\bar{V}_{n,1},\\
A_{t}  &  =A_{0}-t\bar{V}_{n,2},\\
\gamma_{t}  &  =\gamma_{0}-t\bar{V}_{n,0},\\
Q_{t}  &  =Q_{0},\\
P_{t}  &  =P_{0}-t\bar{V}_{n,2}\cdot Q_{0},\\
S_{t}  &  =S_{0}-t\bar{V}_{n,0},
\end{align}
but we still need to evaluate the time averages%
\begin{equation}
\bar{V}_{n,j}:=\frac{1}{t}\int_{0}^{t}V_{n,j}(t^{\prime})dt^{\prime}=\frac
{1}{t}\int_{0}^{t}\mathbf{c}_{t^{\prime}}^{\dag}\mathbf{V}_{j}\mathbf{c}%
_{t^{\prime}}dt^{\prime}%
\end{equation}
of the expectation values of $\mathbf{V}_{j}$ over an evolving electronic
state $\mathbf{c}_{t}$ for $j=0,1,2$. All of the time averages $\bar{V}_{n,j}$
require evaluating the integral%
\begin{equation}
I=\int_{0}^{t}\mathbf{c}_{0}^{\dag}\exp(it^{\prime}\mathbf{V}_{\text{eff}%
}/\hbar)\mathbf{B}\exp(-it^{\prime}\mathbf{V}_{\text{eff}}/\hbar
)\mathbf{c}_{0}dt^{\prime}, \label{eq:I_integral}%
\end{equation}
where the $S\times S$ matrix $\mathbf{B}$ is $\mathbf{V}_{0}$ (for $\bar
{V}_{n,0}$) or one of the $D$ components of vector $\mathbf{V}_{1}$ (for
$\bar{V}_{n,1}$), or one of the $D\times D$ matrix elements of $\mathbf{V}%
_{2}$ (for $\bar{V}_{n,2}$). Using a formula for a derivative of an
exponential with respect to a parameter,\cite{Petersen_Pedersen:2012} one can
recognize that%
\begin{equation}
I=\left.  df\left(  \lambda\right)  /d\lambda\right\vert _{\lambda=0},
\label{eq:I_as_der_of_f}%
\end{equation}
where $f\left(  \lambda\right)  $ is the function%
\begin{equation}
f\left(  \lambda\right)  :=\mathbf{c}_{0}^{\dag}\exp\{t[i\mathbf{V}%
_{\text{eff}}(\psi_{0})\mathbf{/\hbar}+\lambda\mathbf{B]\}c}_{t}.
\label{eq:f_lambda}%
\end{equation}
Equation~(\ref{eq:I_as_der_of_f}) is much more convenient for evaluating $I$
than is the integral in Eq.~(\ref{eq:I_integral}), because it is easier to
evaluate a derivative by finite difference than an integral. E.g., one can
approximate the result by considering small yet finite $\lambda$ in the
expression%
\begin{equation}
I=\lim_{\lambda\rightarrow0}[f(\lambda)-1]/\lambda.
\end{equation}
However, there exist much more accurate, higher-order numerical methods to
estimate the derivative of the function $f(\lambda)$, which have been used for
analogous potential propagation of standard Ehrenfest
dynamics.\cite{Vanicek_Choi:2026}

\section{Exactness of various approximations\label{sec:exactness}}

Let us discuss the conditions, under which various approximations become exact.

\subsection{Time-dependent Hartree approximation\label{sec:exactness_TDH}}

The TDH approximation [Eqs.~(\ref{eq:TDH_mol_ansatz}) and (\ref{eq:TDH_a_t}%
)--(\ref{eq:TDH_mean_field_H_e})]\ is exact if electronic and nuclear dynamics
are independent, which happens when molecular Hamiltonian $\mathcal{H}$ is a
sum%
\begin{equation}
\mathcal{H}=\hat{H}_{\text{nu}}\otimes\hat{1}_{\text{el}}+\hat{1}_{\text{nu}%
}\otimes\hat{H}_{\text{el}}=\hat{H}_{\text{nu}}+\hat{H}_{\text{el}}
\label{eq:H_nonint}%
\end{equation}
of noninteracting nuclear and electronic terms.

\emph{Proof}. The TDH equations (\ref{eq:TDH_a_t}%
)--(\ref{eq:TDH_mean_field_TDSE_e}), which rely on the mean-field Hamiltonians%
\begin{align}
\hat{H}_{n}  &  :=\langle\varphi|\mathcal{H}|\varphi\rangle=\hat{H}%
_{\text{nu}}+\langle\hat{H}_{\text{el}}\rangle_{e},\\
\hat{H}_{e}  &  :=\langle\psi|\mathcal{H}|\psi\rangle=\langle\hat
{H}_{\text{nu}}\rangle_{n}+\hat{H}_{\text{el}},
\end{align}
give%
\begin{align}
i\hbar\dot{\Psi}  &  =i\hbar\lbrack a(\dot{\psi}\varphi+\psi\dot{\varphi
})+\dot{a}\psi\varphi]\nonumber\\
&  =a[(\hat{H}_{n}+\hat{H}_{e})-E]\psi\varphi\nonumber\\
&  =[(\hat{H}_{\text{nu}}+\hat{H}_{\text{el}})+\langle\hat{H}_{\text{nu}}%
+\hat{H}_{\text{el}}\rangle_{\Psi}-E]\Psi\nonumber\\
&  =(\mathcal{H}+E-E)\Psi=\mathcal{H}\Psi,
\end{align}
and therefore solve the molecular TDSE exactly.

\emph{Remark}. Contrary to common intuition, when $\mathcal{H}$ is expressed
in a diabatic basis as $\mathbf{H}(\hat{q},\hat{p})$, the validity of the
TDH\ approximation, requiring form~(\ref{eq:H_nonint}) of $\mathcal{H}$,
\emph{allows} $\mathbf{H}(\hat{q},\hat{p})$ to contain constant couplings
between different electronic states. Also surprisingly, even if $\mathbf{H}%
(\hat{q},\hat{p})$ is a diagonal matrix and contains no offdiagonal coupling
terms, the form~(\ref{eq:H_nonint}) \emph{does not allow} nuclear potential
energy surfaces in the diagonal terms of $\mathbf{H}(\hat{q},\hat{p})$ to
differ more than by a constant vertical shift. Both statements become clear by
expressing Eq.~(\ref{eq:H_nonint}) in diabatic basis:
\begin{equation}
\mathbf{H}(\hat{q},\hat{p})=H_{\text{nu}}(\hat{q},\hat{p})\mathbf{1}+\hat
{1}\mathbf{H}_{\text{el}}. \label{eq:H_nonint_diab_rep}%
\end{equation}
In summary, the TDH\ approximation is exact if all potential energy surfaces
are only vertically displaced and the couplings between these surfaces are constant.

\subsection{Thawed Gaussian wavepacket dynamics\label{sec:exactness_TGWD}}

The single-surface, purely nuclear TGWD from Sec.~\ref{sec:limit_TGWD} is
exact\cite{Heller:1975,book_Tannor:2007,Lasser_Lubich:2020,Vanicek:2023} if
the nuclear potential is a quadratic function%
\begin{equation}
V_{n}(q)=v_{0}+v_{1}^{T}\cdot x_{r}+x_{r}^{T}\cdot v_{2}\cdot x_{r}/2
\label{eq:V_quadratic}%
\end{equation}
of nuclear coordinates.\cite{Vanicek:2023,Lasser_Lubich:2020} As in
Sec.~\ref{sec:H-TGED}, $x_{r}:=q-q_{r}$ is the displacement from a reference
position $q_{r}$, and $v_{j}:=V^{(j)}(q_{r})$ is the $j$th derivative of the
potential at $q_{r}$.

\subsection{Thawed Gaussian Ehrenfest dynamics\label{sec:exactness_TGED}}

The TGED is exact if the molecular Hamiltonian $\mathcal{H}$ has the form of
Eq.~(\ref{eq:H_nonint}), where%
\begin{equation}
\hat{H}_{\text{nu}}=T(\hat{p})+V_{\text{nu}}(\hat{q}),
\end{equation}
and $T(p)$ and $V_{\text{nu}}(q)$ are at most quadratic functions. In diabatic
representation, this means that the potential energy surfaces are vertically
displaced harmonic potentials and the couplings between the surfaces are constant.

\emph{Proof}. In Sec.~\ref{sec:exactness_TDH}, we already showed that the
TDH\ was exact for any Hamiltonian $\mathcal{H}$ of the form of
Eq.~(\ref{eq:H_nonint}). It remains to show that the Gaussian wavepacket
dynamics solves exactly the nuclear equation of the TDH\ approximation. The
condition on $T(p)$ is satisfied because we assume a standard, quadratic
kinetic energy~(\ref{eq:T_p}). If, in addition, $V_{\text{nu}}(q)$ is at most
quadratic, i.e., of the form of the right-hand side of
Eq.~(\ref{eq:V_quadratic}), then the mean-field potential $V_{n}(q)$ will
retain the same form, except that the constant coefficient $v_{0}$ will be
replaced with $v_{0}+E_{\text{el}}$, which is independent of time because%
\[
\langle\hat{H}_{\text{el}}\rangle_{e}=\mathbf{c}_{t}^{\dag}\mathbf{H}%
_{\text{el}}\mathbf{c}_{t}=\mathbf{c}_{0}^{\dag}\mathbf{H}_{\text{el}%
}\mathbf{c}_{0}=E_{\text{el}}%
\]
is the conserved electronic energy. Here, we used the fact that the electronic
and nuclear Hamiltonians commute for $\mathcal{H}$ of Eq.~(\ref{eq:H_nonint}).
Because the mean-field nuclear potential
\begin{equation}
V_{n}(\hat{q})=E_{\text{el}}+V_{\text{nu}}(\hat{q})
\end{equation}
is quadratic of the form~(\ref{eq:V_quadratic}), by
Sec.~\ref{sec:exactness_TGWD} the corresponding nuclear Gaussian wavepacket
dynamics is exact.

\subsection{Ehrenfest dynamics\label{sec:exactness_ED}}

Under the conditions of exactness of the TGED from
Sec.~\ref{sec:exactness_TGED}, the nuclear positions and momenta satisfy
Eqs.~(\ref{eq:q_dot}) and (\ref{eq:p_dot}) with $V_{n,1}=V_{n}^{\prime}%
(q_{t})$, which are already equivalent to the corresponding
Eqs.~(\ref{eq:q_dot_Ehr_diab}) and (\ref{eq:p_dot_Ehr_diab}) of Ehrenfest
dynamics. In the limit $\hbar\rightarrow0$ (for nuclei only!) used in deriving
the Ehrenfest dynamics from a general TGED in Sec.~\ref{sec:limit_Ehrenfest},
both position and momentum widths of the wavepacket vanish and, as a
consequence, the expectation values of kinetic and potential energies converge
to their classical values $T_{n}(p_{t})$ and $V_{n}(q_{t})$ at the center of
the wavepacket [see Eqs.~(\ref{eq:T_GWP_small_h})--(\ref{eq:V_GWP_small_h})].

For nonzero $\hbar$, the equalities%
\begin{equation}
\langle T_{n}(\hat{p})\rangle_{n}=T(p_{t})\text{ \ \ and \ \ }\langle
V_{n}(\hat{q})\rangle_{n}=V_{n}(q_{t})
\end{equation}
will be guaranteed only if both kinetic and potential energies are linear
(rather than quadratic) functions of momenta and positions.\ In other words,
the local linear approximation used for deriving the Ehrenfest dynamics in
Sec.~\ref{sec:limit_Ehrenfest} must be replaced with a global linear kinetic
(this is unusual) and potential energies:%
\begin{align}
T_{n}(p)  &  =t_{0}+t_{1}^{T}\cdot p,\\
V_{n}(q)  &  =v_{0}+v_{1}^{T}\cdot(q-q_{r}).
\end{align}
Of course, these give rather trivial equations of motion%
\begin{equation}
\dot{q}_{t}=t_{1}\text{ \ \ and \ \ }\dot{p}_{t}=-v_{1}%
\end{equation}
with analytical solutions%
\begin{align}
q_{t}  &  =q_{0}+t\,t_{1},\\
p_{t}  &  =p_{0}-t\,v_{1}.
\end{align}
For such a Hamiltonian, the Ehrenfest evolution of even the electronic
wavefunction (\ref{eq:c_dot_Ehr_diab}) is exact since it is equivalent to the
electronic TDSE (\ref{eq:TDSE_el_GWP}) in the TGED.

\section{Conclusion\label{sec:conclusion}}

In conclusion, we have described the single-trajectory thawed Gaussian
Ehrenfest dynamics, which combines the strengths of mixed quantum--classical
Ehrenfest dynamics and semiclassical Gaussian wavepacket dynamics. The method
captures both electronic nonadiabaticity and nuclear quantum effects within a
unified framework while retaining the computational efficiency of a
single-trajectory approach. At the same time, it inherits the principal
limitations of its parent methods: as a mean-field approximation, it cannot
describe wavepacket branching, and, like thawed Gaussian wavepacket dynamics,
it is most accurate for weakly anharmonic systems. These limitations become
particularly apparent near conical intersections. Whereas TGED fails for
conical intersections between electronic states of different symmetry, it
performs surprisingly well for conical intersections involving states of the
same symmetry.\cite{Scheidegger_Vanicek:2026}

The TGED is not constrained to coupled electronic-nuclear dynamics; the
underlying formalism is applicable more generally to systems containing both
quantum and more classical-like degrees of freedom. Owing to its simple
structure, the method can be naturally extended to mixed states by replacing
wavefunctions with density matrices. It may also be systematically improved by
incorporating spin-mapping techniques, analogous to recent advances that have
enhanced Ehrenfest dynamics through spin mapping\cite{Runeson_Richardson:2019}
and nonadiabatic-field methods.\cite{Wu_Liu:2025} Finally, although this work
has focused primarily on population dynamics, the mean-field character of TGED
suggests that, like Ehrenfest dynamics, it should perform better for
electronic coherences, making it a promising tool for the simulations of spectra.

\begin{acknowledgments}
The author acknowledges the financial support from EPFL and thanks Alan
Scheidegger for useful discussions and producing
Figs.~\ref{fig:TGED_comparison} and \ref{fig:TGED}.
\end{acknowledgments}

\section*{Author declarations}

\subsection*{Conflict of interest}

The author has no conflicts to disclose.

\section*{Data availability}

This study did not generate any data.

\bibliographystyle{aipnum4-2}

\begin{thebibliography}{68}%
\makeatletter
\providecommand \@ifxundefined [1]{%
 \@ifx{#1\undefined}
}%
\providecommand \@ifnum [1]{%
 \ifnum #1\expandafter \@firstoftwo
 \else \expandafter \@secondoftwo
 \fi
}%
\providecommand \@ifx [1]{%
 \ifx #1\expandafter \@firstoftwo
 \else \expandafter \@secondoftwo
 \fi
}%
\providecommand \natexlab [1]{#1}%
\providecommand \enquote  [1]{``#1''}%
\providecommand \bibnamefont  [1]{#1}%
\providecommand \bibfnamefont [1]{#1}%
\providecommand \citenamefont [1]{#1}%
\providecommand \href@noop [0]{\@secondoftwo}%
\providecommand \href [0]{\begingroup \@sanitize@url \@href}%
\providecommand \@href[1]{\@@startlink{#1}\@@href}%
\providecommand \@@href[1]{\endgroup#1\@@endlink}%
\providecommand \@sanitize@url [0]{\catcode `\\12\catcode `\$12\catcode
  `\&12\catcode `\#12\catcode `\^12\catcode `\_12\catcode `\%12\relax}%
\providecommand \@@startlink[1]{}%
\providecommand \@@endlink[0]{}%
\providecommand \url  [0]{\begingroup\@sanitize@url \@url }%
\providecommand \@url [1]{\endgroup\@href {#1}{\urlprefix }}%
\providecommand \urlprefix  [0]{URL }%
\providecommand \Eprint [0]{\href }%
\providecommand \doibase [0]{https://doi.org/}%
\providecommand \selectlanguage [0]{\@gobble}%
\providecommand \bibinfo  [0]{\@secondoftwo}%
\providecommand \bibfield  [0]{\@secondoftwo}%
\providecommand \translation [1]{[#1]}%
\providecommand \BibitemOpen [0]{}%
\providecommand \bibitemStop [0]{}%
\providecommand \bibitemNoStop [0]{.\EOS\space}%
\providecommand \EOS [0]{\spacefactor3000\relax}%
\providecommand \BibitemShut  [1]{\csname bibitem#1\endcsname}%
\let\auto@bib@innerbib\@empty
\bibitem [{\citenamefont {Born}\ and\ \citenamefont
  {Oppenheimer}(1927)}]{Born_Oppenheimer:1927}%
  \BibitemOpen
  \bibfield  {author} {\bibinfo {author} {\bibfnamefont {M.}~\bibnamefont
  {Born}}\ and\ \bibinfo {author} {\bibfnamefont {R.}~\bibnamefont
  {Oppenheimer}},\ }\href {https://doi.org/10.1002/andp.19273892002} {\bibfield
   {journal} {\bibinfo  {journal} {Ann.~d.~Phys.}\ }\textbf {\bibinfo {volume}
  {389}},\ \bibinfo {pages} {457} (\bibinfo {year} {1927})}\BibitemShut
  {NoStop}%
\bibitem [{\citenamefont {Tannor}(2007)}]{book_Tannor:2007}%
  \BibitemOpen
  \bibfield  {author} {\bibinfo {author} {\bibfnamefont {D.~J.}\ \bibnamefont
  {Tannor}},\ }\href@noop {} {\emph {\bibinfo {title} {Introduction to Quantum
  Mechanics: A Time-Dependent Perspective}}}\ (\bibinfo  {publisher}
  {University Science Books},\ \bibinfo {address} {Sausalito},\ \bibinfo {year}
  {2007})\BibitemShut {NoStop}%
\bibitem [{\citenamefont {Domcke}\ and\ \citenamefont
  {Yarkony}(2012)}]{Domcke_Yarkony:2012}%
  \BibitemOpen
  \bibfield  {author} {\bibinfo {author} {\bibfnamefont {W.}~\bibnamefont
  {Domcke}}\ and\ \bibinfo {author} {\bibfnamefont {D.~R.}\ \bibnamefont
  {Yarkony}},\ }\href {https://doi.org/10.1146/annurev-physchem-032210-103522}
  {\bibfield  {journal} {\bibinfo  {journal} {Annu.\ Rev.\ Phys.\ Chem.}\
  }\textbf {\bibinfo {volume} {63}},\ \bibinfo {pages} {325} (\bibinfo {year}
  {2012})}\BibitemShut {NoStop}%
\bibitem [{\citenamefont {Agostini}\ and\ \citenamefont
  {Curchod}(2019)}]{Agostini_Curchod:2019}%
  \BibitemOpen
  \bibfield  {author} {\bibinfo {author} {\bibfnamefont {F.}~\bibnamefont
  {Agostini}}\ and\ \bibinfo {author} {\bibfnamefont {B.~F.~E.}\ \bibnamefont
  {Curchod}},\ }\href {https://doi.org/https://doi.org/10.1002/wcms.1417}
  {\bibfield  {journal} {\bibinfo  {journal} {WIREs Comput.\ Mol.\ Sci.}\
  }\textbf {\bibinfo {volume} {9}},\ \bibinfo {pages} {e1417} (\bibinfo {year}
  {2019})}\BibitemShut {NoStop}%
\bibitem [{\citenamefont {Choi}\ and\ \citenamefont
  {Van\'{i}\v{c}ek}(2019)}]{Choi_Vanicek:2019}%
  \BibitemOpen
  \bibfield  {author} {\bibinfo {author} {\bibfnamefont {S.}~\bibnamefont
  {Choi}}\ and\ \bibinfo {author} {\bibfnamefont {J.}~\bibnamefont
  {Van\'{i}\v{c}ek}},\ }\href {https://doi.org/10.1063/1.5092611} {\bibfield
  {journal} {\bibinfo  {journal} {J.~Chem.\ Phys.}\ }\textbf {\bibinfo {volume}
  {150}},\ \bibinfo {pages} {204112} (\bibinfo {year} {2019})}\BibitemShut
  {NoStop}%
\bibitem [{\citenamefont {Roulet}, \citenamefont {Choi},\ and\ \citenamefont
  {Van\'{i}\v{c}ek}(2019)}]{Roulet_Vanicek:2019}%
  \BibitemOpen
  \bibfield  {author} {\bibinfo {author} {\bibfnamefont {J.}~\bibnamefont
  {Roulet}}, \bibinfo {author} {\bibfnamefont {S.}~\bibnamefont {Choi}},\ and\
  \bibinfo {author} {\bibfnamefont {J.}~\bibnamefont {Van\'{i}\v{c}ek}},\
  }\href {https://doi.org/10.1063/1.5094046} {\bibfield  {journal} {\bibinfo
  {journal} {J.~Chem.\ Phys.}\ }\textbf {\bibinfo {volume} {150}},\ \bibinfo
  {pages} {204113} (\bibinfo {year} {2019})}\BibitemShut {NoStop}%
\bibitem [{\citenamefont {Abedi}, \citenamefont {Maitra},\ and\ \citenamefont
  {Gross}(2010)}]{Abedi_Gross:2010}%
  \BibitemOpen
  \bibfield  {author} {\bibinfo {author} {\bibfnamefont {A.}~\bibnamefont
  {Abedi}}, \bibinfo {author} {\bibfnamefont {N.~T.}\ \bibnamefont {Maitra}},\
  and\ \bibinfo {author} {\bibfnamefont {E.~K.}\ \bibnamefont {Gross}},\
  }\href@noop {} {\bibfield  {journal} {\bibinfo  {journal} {Phys.\ Rev.\
  Lett.}\ }\textbf {\bibinfo {volume} {105}},\ \bibinfo {pages} {123002}
  (\bibinfo {year} {2010})}\BibitemShut {NoStop}%
\bibitem [{\citenamefont {Meyer}, \citenamefont {Manthe},\ and\ \citenamefont
  {Cederbaum}(1990)}]{Meyer_Cederbaum:1990}%
  \BibitemOpen
  \bibfield  {author} {\bibinfo {author} {\bibfnamefont {H.-D.}\ \bibnamefont
  {Meyer}}, \bibinfo {author} {\bibfnamefont {U.}~\bibnamefont {Manthe}},\ and\
  \bibinfo {author} {\bibfnamefont {L.~S.}\ \bibnamefont {Cederbaum}},\ }\href
  {https://doi.org/10.1016/0009-2614(90)87014-I} {\bibfield  {journal}
  {\bibinfo  {journal} {Chem.\ Phys.\ Lett.}\ }\textbf {\bibinfo {volume}
  {165}},\ \bibinfo {pages} {73} (\bibinfo {year} {1990})}\BibitemShut
  {NoStop}%
\bibitem [{\citenamefont {Mart\'{\i}nez}\ and\ \citenamefont
  {Levine}(1997)}]{Martinez_Levine:1997}%
  \BibitemOpen
  \bibfield  {author} {\bibinfo {author} {\bibfnamefont {T.~J.}\ \bibnamefont
  {Mart\'{\i}nez}}\ and\ \bibinfo {author} {\bibfnamefont {R.~D.}\ \bibnamefont
  {Levine}},\ }\href {https://doi.org/10.1039/A605958I} {\bibfield  {journal}
  {\bibinfo  {journal} {J.~Chem.\ Soc., Faraday Trans.}\ }\textbf {\bibinfo
  {volume} {93}},\ \bibinfo {pages} {941} (\bibinfo {year} {1997})}\BibitemShut
  {NoStop}%
\bibitem [{\citenamefont {Worth}, \citenamefont {Robb},\ and\ \citenamefont
  {Burghardt}(2004)}]{Worth_Burghardt:2004}%
  \BibitemOpen
  \bibfield  {author} {\bibinfo {author} {\bibfnamefont {G.~A.}\ \bibnamefont
  {Worth}}, \bibinfo {author} {\bibfnamefont {M.~A.}\ \bibnamefont {Robb}},\
  and\ \bibinfo {author} {\bibfnamefont {I.}~\bibnamefont {Burghardt}},\ }\href
  {https://doi.org/10.1039/B314253A} {\bibfield  {journal} {\bibinfo  {journal}
  {Faraday Discuss.}\ }\textbf {\bibinfo {volume} {127}},\ \bibinfo {pages}
  {307} (\bibinfo {year} {2004})}\BibitemShut {NoStop}%
\bibitem [{\citenamefont {Shalashilin}(2009)}]{Shalashilin:2009}%
  \BibitemOpen
  \bibfield  {author} {\bibinfo {author} {\bibfnamefont {D.~V.}\ \bibnamefont
  {Shalashilin}},\ }\href {https://doi.org/10.1063/1.3153302} {\bibfield
  {journal} {\bibinfo  {journal} {J.~Chem.\ Phys.}\ }\textbf {\bibinfo {volume}
  {130}},\ \bibinfo {pages} {244101} (\bibinfo {year} {2009})}\BibitemShut
  {NoStop}%
\bibitem [{\citenamefont {Kapral}\ and\ \citenamefont
  {Ciccotti}(1999)}]{Kapral_Ciccotti:1999}%
  \BibitemOpen
  \bibfield  {author} {\bibinfo {author} {\bibfnamefont {R.}~\bibnamefont
  {Kapral}}\ and\ \bibinfo {author} {\bibfnamefont {G.}~\bibnamefont
  {Ciccotti}},\ }\href {https://doi.org/10.1063/1.478811} {\bibfield  {journal}
  {\bibinfo  {journal} {J.~Chem.\ Phys.}\ }\textbf {\bibinfo {volume} {110}},\
  \bibinfo {pages} {8919} (\bibinfo {year} {1999})}\BibitemShut {NoStop}%
\bibitem [{\citenamefont {Huo}\ and\ \citenamefont
  {Coker}(2011)}]{Huo_Coker:2011}%
  \BibitemOpen
  \bibfield  {author} {\bibinfo {author} {\bibfnamefont {P.}~\bibnamefont
  {Huo}}\ and\ \bibinfo {author} {\bibfnamefont {D.~F.}\ \bibnamefont
  {Coker}},\ }\href {https://doi.org/10.1063/1.3664763} {\bibfield  {journal}
  {\bibinfo  {journal} {J.~Chem.\ Phys.}\ }\textbf {\bibinfo {volume} {135}},\
  \bibinfo {eid} {201101} (\bibinfo {year} {2011})}\BibitemShut {NoStop}%
\bibitem [{\citenamefont {Runeson}\ and\ \citenamefont
  {Richardson}(2020)}]{Runeson_Richardson:2020}%
  \BibitemOpen
  \bibfield  {author} {\bibinfo {author} {\bibfnamefont {J.~E.}\ \bibnamefont
  {Runeson}}\ and\ \bibinfo {author} {\bibfnamefont {J.~O.}\ \bibnamefont
  {Richardson}},\ }\href {https://doi.org/10.1063/1.5143412} {\bibfield
  {journal} {\bibinfo  {journal} {J.~Chem.\ Phys.}\ }\textbf {\bibinfo {volume}
  {152}},\ \bibinfo {pages} {084110} (\bibinfo {year} {2020})}\BibitemShut
  {NoStop}%
\bibitem [{\citenamefont {Wu}\ \emph {et~al.}(2025)\citenamefont {Wu},
  \citenamefont {Li}, \citenamefont {He}, \citenamefont {Cheng}, \citenamefont
  {Ren},\ and\ \citenamefont {Liu}}]{Wu_Liu:2025}%
  \BibitemOpen
  \bibfield  {author} {\bibinfo {author} {\bibfnamefont {B.}~\bibnamefont
  {Wu}}, \bibinfo {author} {\bibfnamefont {B.}~\bibnamefont {Li}}, \bibinfo
  {author} {\bibfnamefont {X.}~\bibnamefont {He}}, \bibinfo {author}
  {\bibfnamefont {X.}~\bibnamefont {Cheng}}, \bibinfo {author} {\bibfnamefont
  {J.}~\bibnamefont {Ren}},\ and\ \bibinfo {author} {\bibfnamefont
  {J.}~\bibnamefont {Liu}},\ }\href {https://doi.org/10.1021/acs.jctc.5c00181}
  {\bibfield  {journal} {\bibinfo  {journal} {J.~Chem.\ Theory Comput.}\
  }\textbf {\bibinfo {volume} {21}},\ \bibinfo {pages} {3775} (\bibinfo {year}
  {2025})}\BibitemShut {NoStop}%
\bibitem [{\citenamefont {Belyaev}, \citenamefont {Lasser},\ and\ \citenamefont
  {Trigila}(2014)}]{Belyaev_Trigila:2014}%
  \BibitemOpen
  \bibfield  {author} {\bibinfo {author} {\bibfnamefont {A.~K.}\ \bibnamefont
  {Belyaev}}, \bibinfo {author} {\bibfnamefont {C.}~\bibnamefont {Lasser}},\
  and\ \bibinfo {author} {\bibfnamefont {G.}~\bibnamefont {Trigila}},\ }\href
  {https://doi.org/10.1063/1.4882073} {\bibfield  {journal} {\bibinfo
  {journal} {J.~Chem.\ Phys.}\ }\textbf {\bibinfo {volume} {140}},\ \bibinfo
  {pages} {224108} (\bibinfo {year} {2014})}\BibitemShut {NoStop}%
\bibitem [{\citenamefont {Subotnik}\ \emph {et~al.}(2016)\citenamefont
  {Subotnik}, \citenamefont {Jain}, \citenamefont {Landry}, \citenamefont
  {Petit}, \citenamefont {Ouyang},\ and\ \citenamefont
  {Bellonzi}}]{Subotnik_Bellonzi:2016}%
  \BibitemOpen
  \bibfield  {author} {\bibinfo {author} {\bibfnamefont {J.~E.}\ \bibnamefont
  {Subotnik}}, \bibinfo {author} {\bibfnamefont {A.}~\bibnamefont {Jain}},
  \bibinfo {author} {\bibfnamefont {B.}~\bibnamefont {Landry}}, \bibinfo
  {author} {\bibfnamefont {A.}~\bibnamefont {Petit}}, \bibinfo {author}
  {\bibfnamefont {W.}~\bibnamefont {Ouyang}},\ and\ \bibinfo {author}
  {\bibfnamefont {N.}~\bibnamefont {Bellonzi}},\ }\href
  {https://doi.org/10.1146/annurev-physchem-040215-112245} {\bibfield
  {journal} {\bibinfo  {journal} {Annu.\ Rev.\ Phys.\ Chem.}\ }\textbf
  {\bibinfo {volume} {67}},\ \bibinfo {pages} {387} (\bibinfo {year}
  {2016})}\BibitemShut {NoStop}%
\bibitem [{\citenamefont {Mannouch}\ and\ \citenamefont
  {Richardson}(2023)}]{Mannouch_Richardson:2023}%
  \BibitemOpen
  \bibfield  {author} {\bibinfo {author} {\bibfnamefont {J.~R.}\ \bibnamefont
  {Mannouch}}\ and\ \bibinfo {author} {\bibfnamefont {J.~O.}\ \bibnamefont
  {Richardson}},\ }\href {https://doi.org/10.1063/5.0139734} {\bibfield
  {journal} {\bibinfo  {journal} {J.~Chem.\ Phys.}\ }\textbf {\bibinfo {volume}
  {158}},\ \bibinfo {pages} {104111} (\bibinfo {year} {2023})}\BibitemShut
  {NoStop}%
\bibitem [{\citenamefont {Tully}(1990)}]{Tully:1990}%
  \BibitemOpen
  \bibfield  {author} {\bibinfo {author} {\bibfnamefont {J.~C.}\ \bibnamefont
  {Tully}},\ }\href {https://doi.org/10.1063/1.459170} {\bibfield  {journal}
  {\bibinfo  {journal} {J.~Chem.\ Phys.}\ }\textbf {\bibinfo {volume} {93}},\
  \bibinfo {pages} {1061} (\bibinfo {year} {1990})}\BibitemShut {NoStop}%
\bibitem [{\citenamefont {Ehrenfest}(1927)}]{Ehrenfest:1927}%
  \BibitemOpen
  \bibfield  {author} {\bibinfo {author} {\bibfnamefont {P.}~\bibnamefont
  {Ehrenfest}},\ }\href@noop {} {\bibfield  {journal} {\bibinfo  {journal} {Z.
  Phys}\ }\textbf {\bibinfo {volume} {45}},\ \bibinfo {pages} {455} (\bibinfo
  {year} {1927})}\BibitemShut {NoStop}%
\bibitem [{\citenamefont {Tully}(1998)}]{Tully:1998}%
  \BibitemOpen
  \bibfield  {author} {\bibinfo {author} {\bibfnamefont {J.~C.}\ \bibnamefont
  {Tully}},\ }\href@noop {} {\bibfield  {journal} {\bibinfo  {journal} {Faraday
  Discuss.}\ }\textbf {\bibinfo {volume} {110}},\ \bibinfo {pages} {407}
  (\bibinfo {year} {1998})}\BibitemShut {NoStop}%
\bibitem [{\citenamefont {Scheidegger}\ and\ \citenamefont
  {Van\'{\i}\v{c}ek}(2025{\natexlab{a}})}]{Scheidegger_Vanicek:2025a}%
  \BibitemOpen
  \bibfield  {author} {\bibinfo {author} {\bibfnamefont {A.}~\bibnamefont
  {Scheidegger}}\ and\ \bibinfo {author} {\bibfnamefont {J.~J.~L.}\
  \bibnamefont {Van\'{\i}\v{c}ek}},\ }\href {https://doi.org/10.1063/5.0276025}
  {\bibfield  {journal} {\bibinfo  {journal} {J.~Chem.\ Phys.}\ }\textbf
  {\bibinfo {volume} {163}},\ \bibinfo {pages} {044105} (\bibinfo {year}
  {2025}{\natexlab{a}})}\BibitemShut {NoStop}%
\bibitem [{\citenamefont {Zeiri}\ and\ \citenamefont
  {Kosloff}(1990)}]{Zeiri_Kosloff:1990}%
  \BibitemOpen
  \bibfield  {author} {\bibinfo {author} {\bibfnamefont {Y.}~\bibnamefont
  {Zeiri}}\ and\ \bibinfo {author} {\bibfnamefont {R.}~\bibnamefont
  {Kosloff}},\ }\href {https://doi.org/10.1063/1.458922} {\bibfield  {journal}
  {\bibinfo  {journal} {J.~Chem.\ Phys.}\ }\textbf {\bibinfo {volume} {93}},\
  \bibinfo {pages} {6890} (\bibinfo {year} {1990})}\BibitemShut {NoStop}%
\bibitem [{\citenamefont {Akimov}, \citenamefont {Long},\ and\ \citenamefont
  {Prezhdo}(2014)}]{Akimov_Prezhdo:2014}%
  \BibitemOpen
  \bibfield  {author} {\bibinfo {author} {\bibfnamefont {A.~V.}\ \bibnamefont
  {Akimov}}, \bibinfo {author} {\bibfnamefont {R.}~\bibnamefont {Long}},\ and\
  \bibinfo {author} {\bibfnamefont {O.~V.}\ \bibnamefont {Prezhdo}},\ }\href
  {https://doi.org/10.1063/1.4875702} {\bibfield  {journal} {\bibinfo
  {journal} {J.~Chem.\ Phys.}\ }\textbf {\bibinfo {volume} {140}},\ \bibinfo
  {pages} {194107} (\bibinfo {year} {2014})}\BibitemShut {NoStop}%
\bibitem [{\citenamefont {Gherib}, \citenamefont {Ryabinkin},\ and\
  \citenamefont {Izmaylov}(2015)}]{Gherib_Izmaylov:2015}%
  \BibitemOpen
  \bibfield  {author} {\bibinfo {author} {\bibfnamefont {R.}~\bibnamefont
  {Gherib}}, \bibinfo {author} {\bibfnamefont {I.~G.}\ \bibnamefont
  {Ryabinkin}},\ and\ \bibinfo {author} {\bibfnamefont {A.~F.}\ \bibnamefont
  {Izmaylov}},\ }\href {https://doi.org/10.1021/acs.jctc.5b00072} {\bibfield
  {journal} {\bibinfo  {journal} {J.~Chem.\ Theory Comput.}\ }\textbf {\bibinfo
  {volume} {11}},\ \bibinfo {pages} {1375} (\bibinfo {year}
  {2015})}\BibitemShut {NoStop}%
\bibitem [{\citenamefont {Miller}(2001)}]{Miller:2001}%
  \BibitemOpen
  \bibfield  {author} {\bibinfo {author} {\bibfnamefont {W.~H.}\ \bibnamefont
  {Miller}},\ }\href {https://doi.org/10.1021/jp003712k} {\bibfield  {journal}
  {\bibinfo  {journal} {J.~Phys.\ Chem.~A}\ }\textbf {\bibinfo {volume}
  {105}},\ \bibinfo {pages} {2942} (\bibinfo {year} {2001})}\BibitemShut
  {NoStop}%
\bibitem [{\citenamefont {Vleck}(1928)}]{Vleck:1928}%
  \BibitemOpen
  \bibfield  {author} {\bibinfo {author} {\bibfnamefont {J.~H.~V.}\
  \bibnamefont {Vleck}},\ }\href@noop {} {\bibfield  {journal} {\bibinfo
  {journal} {Proc.\ Nat.\ Acad.\ Sci.\ USA}\ }\textbf {\bibinfo {volume}
  {14}},\ \bibinfo {pages} {178} (\bibinfo {year} {1928})}\BibitemShut
  {NoStop}%
\bibitem [{\citenamefont {Herman}\ and\ \citenamefont
  {Kluk}(1984)}]{Herman_Kluk:1984}%
  \BibitemOpen
  \bibfield  {author} {\bibinfo {author} {\bibfnamefont {M.~F.}\ \bibnamefont
  {Herman}}\ and\ \bibinfo {author} {\bibfnamefont {E.}~\bibnamefont {Kluk}},\
  }\href {https://doi.org/10.1016/0301-0104(84)80039-7} {\bibfield  {journal}
  {\bibinfo  {journal} {Chem.\ Phys.}\ }\textbf {\bibinfo {volume} {91}},\
  \bibinfo {pages} {27} (\bibinfo {year} {1984})}\BibitemShut {NoStop}%
\bibitem [{\citenamefont {Ceotto}, \citenamefont {Di~Liberto},\ and\
  \citenamefont {Conte}(2017)}]{Ceotto_Conte:2017}%
  \BibitemOpen
  \bibfield  {author} {\bibinfo {author} {\bibfnamefont {M.}~\bibnamefont
  {Ceotto}}, \bibinfo {author} {\bibfnamefont {G.}~\bibnamefont {Di~Liberto}},\
  and\ \bibinfo {author} {\bibfnamefont {R.}~\bibnamefont {Conte}},\ }\href
  {https://doi.org/10.1103/PhysRevLett.119.010401} {\bibfield  {journal}
  {\bibinfo  {journal} {Phys.\ Rev.\ Lett.}\ }\textbf {\bibinfo {volume}
  {119}},\ \bibinfo {pages} {010401} (\bibinfo {year} {2017})}\BibitemShut
  {NoStop}%
\bibitem [{\citenamefont {Heller}(1975)}]{Heller:1975}%
  \BibitemOpen
  \bibfield  {author} {\bibinfo {author} {\bibfnamefont {E.~J.}\ \bibnamefont
  {Heller}},\ }\href {https://doi.org/10.1063/1.430620} {\bibfield  {journal}
  {\bibinfo  {journal} {J.~Chem.\ Phys.}\ }\textbf {\bibinfo {volume} {62}},\
  \bibinfo {pages} {1544} (\bibinfo {year} {1975})}\BibitemShut {NoStop}%
\bibitem [{\citenamefont {Heller}(2018)}]{book_Heller:2018}%
  \BibitemOpen
  \bibfield  {author} {\bibinfo {author} {\bibfnamefont {E.~J.}\ \bibnamefont
  {Heller}},\ }\href@noop {} {\emph {\bibinfo {title} {The semiclassical way to
  dynamics and spectroscopy}}}\ (\bibinfo  {publisher} {Princeton University
  Press},\ \bibinfo {address} {Princeton, NJ},\ \bibinfo {year}
  {2018})\BibitemShut {NoStop}%
\bibitem [{\citenamefont {Coalson}\ and\ \citenamefont
  {Karplus}(1990)}]{Coalson_Karplus:1990}%
  \BibitemOpen
  \bibfield  {author} {\bibinfo {author} {\bibfnamefont {R.~D.}\ \bibnamefont
  {Coalson}}\ and\ \bibinfo {author} {\bibfnamefont {M.}~\bibnamefont
  {Karplus}},\ }\href {https://doi.org/10.1063/1.458778} {\bibfield  {journal}
  {\bibinfo  {journal} {J.~Chem.\ Phys.}\ }\textbf {\bibinfo {volume} {93}},\
  \bibinfo {pages} {3919} (\bibinfo {year} {1990})}\BibitemShut {NoStop}%
\bibitem [{\citenamefont {Begu\v{s}i\'{c}}, \citenamefont {Cordova},\ and\
  \citenamefont {Van{\'{i}}{\v{c}}ek}(2019)}]{Begusic_Vanicek:2019}%
  \BibitemOpen
  \bibfield  {author} {\bibinfo {author} {\bibfnamefont {T.}~\bibnamefont
  {Begu\v{s}i\'{c}}}, \bibinfo {author} {\bibfnamefont {M.}~\bibnamefont
  {Cordova}},\ and\ \bibinfo {author} {\bibfnamefont {J.}~\bibnamefont
  {Van{\'{i}}{\v{c}}ek}},\ }\href {https://doi.org/10.1063/1.5090122}
  {\bibfield  {journal} {\bibinfo  {journal} {J.~Chem.\ Phys.}\ }\textbf
  {\bibinfo {volume} {150}},\ \bibinfo {pages} {154117} (\bibinfo {year}
  {2019})}\BibitemShut {NoStop}%
\bibitem [{\citenamefont {Van\'i\v{c}ek}(2023)}]{Vanicek:2023}%
  \BibitemOpen
  \bibfield  {author} {\bibinfo {author} {\bibfnamefont {J.~J.~L.}\
  \bibnamefont {Van\'i\v{c}ek}},\ }\href {https://doi.org/10.1063/5.0146680}
  {\bibfield  {journal} {\bibinfo  {journal} {J.~Chem.\ Phys.}\ }\textbf
  {\bibinfo {volume} {159}},\ \bibinfo {pages} {014114} (\bibinfo {year}
  {2023})}\BibitemShut {NoStop}%
\bibitem [{\citenamefont {Burkhard}\ \emph {et~al.}(2024)\citenamefont
  {Burkhard}, \citenamefont {D{\"o}rich}, \citenamefont {Hochbruck},\ and\
  \citenamefont {Lasser}}]{Burkhard_Lasser:2024}%
  \BibitemOpen
  \bibfield  {author} {\bibinfo {author} {\bibfnamefont {S.}~\bibnamefont
  {Burkhard}}, \bibinfo {author} {\bibfnamefont {B.}~\bibnamefont
  {D{\"o}rich}}, \bibinfo {author} {\bibfnamefont {M.}~\bibnamefont
  {Hochbruck}},\ and\ \bibinfo {author} {\bibfnamefont {C.}~\bibnamefont
  {Lasser}},\ }\href {https://doi.org/10.1088/1751-8121/ad591e} {\bibfield
  {journal} {\bibinfo  {journal} {Journal of Physics A: Mathematical and
  Theoretical}\ }\textbf {\bibinfo {volume} {57}},\ \bibinfo {pages} {295202}
  (\bibinfo {year} {2024})}\BibitemShut {NoStop}%
\bibitem [{\citenamefont {Wehrle}, \citenamefont {\v{S}ulc},\ and\
  \citenamefont {Van\'{i}\v{c}ek}(2014)}]{Wehrle_Vanicek:2014}%
  \BibitemOpen
  \bibfield  {author} {\bibinfo {author} {\bibfnamefont {M.}~\bibnamefont
  {Wehrle}}, \bibinfo {author} {\bibfnamefont {M.}~\bibnamefont {\v{S}ulc}},\
  and\ \bibinfo {author} {\bibfnamefont {J.}~\bibnamefont {Van\'{i}\v{c}ek}},\
  }\href {https://doi.org/10.1063/1.4884718} {\bibfield  {journal} {\bibinfo
  {journal} {J.~Chem.\ Phys.}\ }\textbf {\bibinfo {volume} {140}},\ \bibinfo
  {pages} {244114} (\bibinfo {year} {2014})}\BibitemShut {NoStop}%
\bibitem [{\citenamefont {Kl\={e}tnieks}, \citenamefont {Alonso},\ and\
  \citenamefont {Van\'i\v{c}ek}(2023)}]{Kletnieks_Vanicek:2023}%
  \BibitemOpen
  \bibfield  {author} {\bibinfo {author} {\bibfnamefont {E.}~\bibnamefont
  {Kl\={e}tnieks}}, \bibinfo {author} {\bibfnamefont {Y.~C.}\ \bibnamefont
  {Alonso}},\ and\ \bibinfo {author} {\bibfnamefont {J.~J.~L.}\ \bibnamefont
  {Van\'i\v{c}ek}},\ }\href {https://doi.org/10.1021/acs.jpca.3c04607}
  {\bibfield  {journal} {\bibinfo  {journal} {J.~Phys.\ Chem.~A}\ }\textbf
  {\bibinfo {volume} {127}},\ \bibinfo {pages} {8117} (\bibinfo {year}
  {2023})}\BibitemShut {NoStop}%
\bibitem [{\citenamefont {Begu{\v{s}}i{\'{c}}}\ and\ \citenamefont
  {Van{\'{i}}{\v{c}}ek}(2020)}]{Begusic_Vanicek:2020}%
  \BibitemOpen
  \bibfield  {author} {\bibinfo {author} {\bibfnamefont {T.}~\bibnamefont
  {Begu{\v{s}}i{\'{c}}}}\ and\ \bibinfo {author} {\bibfnamefont
  {J.}~\bibnamefont {Van{\'{i}}{\v{c}}ek}},\ }\href
  {https://doi.org/10.1063/5.0013677} {\bibfield  {journal} {\bibinfo
  {journal} {J.~Chem.\ Phys.}\ }\textbf {\bibinfo {volume} {153}},\ \bibinfo
  {pages} {024105} (\bibinfo {year} {2020})}\BibitemShut {NoStop}%
\bibitem [{\citenamefont {Begu{\v{s}}i{\'{c}}}\ and\ \citenamefont
  {Van{\'{i}}{\v{c}}ek}(2021)}]{Begusic_Vanicek:2021}%
  \BibitemOpen
  \bibfield  {author} {\bibinfo {author} {\bibfnamefont {T.}~\bibnamefont
  {Begu{\v{s}}i{\'{c}}}}\ and\ \bibinfo {author} {\bibfnamefont
  {J.}~\bibnamefont {Van{\'{i}}{\v{c}}ek}},\ }\href
  {https://doi.org/10.1021/acs.jpclett.1c00123} {\bibfield  {journal} {\bibinfo
   {journal} {J.~Phys.\ Chem.\ Lett.}\ }\textbf {\bibinfo {volume} {12}},\
  \bibinfo {pages} {2997} (\bibinfo {year} {2021})}\BibitemShut {NoStop}%
\bibitem [{\citenamefont {Golubev}, \citenamefont {Begu{\v{s}}i{\'{c}}},\ and\
  \citenamefont {Van{\'{i}}{\v{c}}ek}(2020)}]{Golubev_Vanicek:2020}%
  \BibitemOpen
  \bibfield  {author} {\bibinfo {author} {\bibfnamefont {N.~V.}\ \bibnamefont
  {Golubev}}, \bibinfo {author} {\bibfnamefont {T.}~\bibnamefont
  {Begu{\v{s}}i{\'{c}}}},\ and\ \bibinfo {author} {\bibfnamefont
  {J.}~\bibnamefont {Van{\'{i}}{\v{c}}ek}},\ }\href
  {https://doi.org/10.1103/physrevlett.125.083001} {\bibfield  {journal}
  {\bibinfo  {journal} {Phys.\ Rev.\ Lett.}\ }\textbf {\bibinfo {volume}
  {125}},\ \bibinfo {pages} {083001} (\bibinfo {year} {2020})}\BibitemShut
  {NoStop}%
\bibitem [{\citenamefont {Scheidegger}, \citenamefont {Van{\'{i}}{\v{c}}ek},\
  and\ \citenamefont {Golubev}(2022)}]{Scheidegger_Golubev:2022}%
  \BibitemOpen
  \bibfield  {author} {\bibinfo {author} {\bibfnamefont {A.}~\bibnamefont
  {Scheidegger}}, \bibinfo {author} {\bibfnamefont {J.}~\bibnamefont
  {Van{\'{i}}{\v{c}}ek}},\ and\ \bibinfo {author} {\bibfnamefont {N.~V.}\
  \bibnamefont {Golubev}},\ }\href {https://doi.org/10.1063/5.0076609}
  {\bibfield  {journal} {\bibinfo  {journal} {J.~Chem.\ Phys.}\ }\textbf
  {\bibinfo {volume} {156}},\ \bibinfo {pages} {034104} (\bibinfo {year}
  {2022})}\BibitemShut {NoStop}%
\bibitem [{\citenamefont {Scheidegger}, \citenamefont {Golubev},\ and\
  \citenamefont {Van\'{\i}\v{c}ek}(2025)}]{Scheidegger_Vanicek:2025}%
  \BibitemOpen
  \bibfield  {author} {\bibinfo {author} {\bibfnamefont {A.}~\bibnamefont
  {Scheidegger}}, \bibinfo {author} {\bibfnamefont {N.~V.}\ \bibnamefont
  {Golubev}},\ and\ \bibinfo {author} {\bibfnamefont {J.~J.~L.}\ \bibnamefont
  {Van\'{\i}\v{c}ek}},\ }\href {https://doi.org/10.1073/pnas.2501319122}
  {\bibfield  {journal} {\bibinfo  {journal} {Proc.\ Nat.\ Acad.\ Sci.\ USA}\
  }\textbf {\bibinfo {volume} {122}},\ \bibinfo {pages} {e2501319122} (\bibinfo
  {year} {2025})}\BibitemShut {NoStop}%
\bibitem [{\citenamefont {Scheidegger}\ and\ \citenamefont
  {Van\'{\i}\v{c}ek}(2025{\natexlab{b}})}]{Scheidegger_Vanicek:2026}%
  \BibitemOpen
  \bibfield  {author} {\bibinfo {author} {\bibfnamefont {A.}~\bibnamefont
  {Scheidegger}}\ and\ \bibinfo {author} {\bibfnamefont {J.~J.~L.}\
  \bibnamefont {Van\'{\i}\v{c}ek}},\ }\href {http://arxiv.org/abs/2504.05922}
  {\enquote {\bibinfo {title} {{Thawed Gaussian Ehrenfest dynamics at conical
  intersections: When can a single mean-field trajectory capture internal
  conversion?}}}\ } (\bibinfo {year} {2025}{\natexlab{b}}),\ \Eprint
  {https://arxiv.org/abs/2504.05922} {arXiv:2504.05922 [physics.chem-ph]}
  \BibitemShut {NoStop}%
\bibitem [{\citenamefont {Dirac}(1930)}]{Dirac:1930}%
  \BibitemOpen
  \bibfield  {author} {\bibinfo {author} {\bibfnamefont {P.~A.~M.}\
  \bibnamefont {Dirac}},\ }\href {https://doi.org/10.1017/S0305004100016108}
  {\bibfield  {journal} {\bibinfo  {journal} {Math.\ Proc.\ Camb.\ Phil.\
  Soc.}\ }\textbf {\bibinfo {volume} {26}},\ \bibinfo {pages} {376} (\bibinfo
  {year} {1930})}\BibitemShut {NoStop}%
\bibitem [{\citenamefont {Frenkel}(1934)}]{book_Frenkel:1934}%
  \BibitemOpen
  \bibfield  {author} {\bibinfo {author} {\bibfnamefont {J.}~\bibnamefont
  {Frenkel}},\ }\href@noop {} {\emph {\bibinfo {title} {Wave mechanics}}}\
  (\bibinfo  {publisher} {Clarendon Press},\ \bibinfo {address} {Oxford},\
  \bibinfo {year} {1934})\BibitemShut {NoStop}%
\bibitem [{\citenamefont {Lubich}(2008)}]{book_Lubich:2008}%
  \BibitemOpen
  \bibfield  {author} {\bibinfo {author} {\bibfnamefont {C.}~\bibnamefont
  {Lubich}},\ }\href@noop {} {\emph {\bibinfo {title} {From Quantum to
  Classical Molecular Dynamics: Reduced Models and Numerical Analysis}}},\
  \bibinfo {edition} {12th}\ ed.\ (\bibinfo  {publisher} {European Mathematical
  Society},\ \bibinfo {address} {Z\"{u}rich},\ \bibinfo {year}
  {2008})\BibitemShut {NoStop}%
\bibitem [{\citenamefont {Lasser}\ and\ \citenamefont
  {Lubich}(2020)}]{Lasser_Lubich:2020}%
  \BibitemOpen
  \bibfield  {author} {\bibinfo {author} {\bibfnamefont {C.}~\bibnamefont
  {Lasser}}\ and\ \bibinfo {author} {\bibfnamefont {C.}~\bibnamefont
  {Lubich}},\ }\href {https://doi.org/10.1017/S0962492920000033} {\bibfield
  {journal} {\bibinfo  {journal} {Acta\ Numer.}\ }\textbf {\bibinfo {volume}
  {29}},\ \bibinfo {pages} {229} (\bibinfo {year} {2020})}\BibitemShut
  {NoStop}%
\bibitem [{\citenamefont {Choi}\ and\ \citenamefont
  {Van{\'\i}{\v{c}}ek}(2020)}]{Choi_Vanicek:2020}%
  \BibitemOpen
  \bibfield  {author} {\bibinfo {author} {\bibfnamefont {S.}~\bibnamefont
  {Choi}}\ and\ \bibinfo {author} {\bibfnamefont {J.}~\bibnamefont
  {Van{\'\i}{\v{c}}ek}},\ }\href {https://doi.org/10.1063/5.0033410} {\bibfield
   {journal} {\bibinfo  {journal} {J.~Chem.\ Phys.}\ }\textbf {\bibinfo
  {volume} {153}},\ \bibinfo {pages} {211101} (\bibinfo {year}
  {2020})}\BibitemShut {NoStop}%
\bibitem [{\citenamefont {Choi}\ and\ \citenamefont
  {Van\'{i}\v{c}ek}(2021{\natexlab{a}})}]{Choi_Vanicek:2021}%
  \BibitemOpen
  \bibfield  {author} {\bibinfo {author} {\bibfnamefont {S.}~\bibnamefont
  {Choi}}\ and\ \bibinfo {author} {\bibfnamefont {J.}~\bibnamefont
  {Van\'{i}\v{c}ek}},\ }\href {https://doi.org/10.1063/5.0046067} {\bibfield
  {journal} {\bibinfo  {journal} {J.~Chem.\ Phys.}\ }\textbf {\bibinfo {volume}
  {154}},\ \bibinfo {pages} {124119} (\bibinfo {year}
  {2021}{\natexlab{a}})}\BibitemShut {NoStop}%
\bibitem [{\citenamefont {Heller}(1976{\natexlab{a}})}]{Heller:1976}%
  \BibitemOpen
  \bibfield  {author} {\bibinfo {author} {\bibfnamefont {E.~J.}\ \bibnamefont
  {Heller}},\ }\href {https://doi.org/10.1063/1.431911} {\bibfield  {journal}
  {\bibinfo  {journal} {J.~Chem.\ Phys.}\ }\textbf {\bibinfo {volume} {64}},\
  \bibinfo {pages} {63} (\bibinfo {year} {1976}{\natexlab{a}})}\BibitemShut
  {NoStop}%
\bibitem [{\citenamefont {Patoz}, \citenamefont {Begu\v{s}i{\'{c}}},\ and\
  \citenamefont {Van{\'{i}}{\v{c}}ek}(2018)}]{Patoz_Vanicek:2018}%
  \BibitemOpen
  \bibfield  {author} {\bibinfo {author} {\bibfnamefont {A.}~\bibnamefont
  {Patoz}}, \bibinfo {author} {\bibfnamefont {T.}~\bibnamefont
  {Begu\v{s}i{\'{c}}}},\ and\ \bibinfo {author} {\bibfnamefont
  {J.}~\bibnamefont {Van{\'{i}}{\v{c}}ek}},\ }\href
  {https://doi.org/10.1021/acs.jpclett.8b00827} {\bibfield  {journal} {\bibinfo
   {journal} {J.~Phys.\ Chem.\ Lett.}\ }\textbf {\bibinfo {volume} {9}},\
  \bibinfo {pages} {2367} (\bibinfo {year} {2018})}\BibitemShut {NoStop}%
\bibitem [{\citenamefont {Heller}(1976{\natexlab{b}})}]{Heller:1976a}%
  \BibitemOpen
  \bibfield  {author} {\bibinfo {author} {\bibfnamefont {E.~J.}\ \bibnamefont
  {Heller}},\ }\href {https://doi.org/10.1063/1.432974} {\bibfield  {journal}
  {\bibinfo  {journal} {J.~Chem.\ Phys.}\ }\textbf {\bibinfo {volume} {65}},\
  \bibinfo {pages} {4979} (\bibinfo {year} {1976}{\natexlab{b}})}\BibitemShut
  {NoStop}%
\bibitem [{\citenamefont {Hagedorn}(1980)}]{Hagedorn:1980}%
  \BibitemOpen
  \bibfield  {author} {\bibinfo {author} {\bibfnamefont {G.~A.}\ \bibnamefont
  {Hagedorn}},\ }\href {https://doi.org/10.1007/BF01230088} {\bibfield
  {journal} {\bibinfo  {journal} {Commun.\ Math.\ Phys.}\ }\textbf {\bibinfo
  {volume} {71}},\ \bibinfo {pages} {77} (\bibinfo {year} {1980})}\BibitemShut
  {NoStop}%
\bibitem [{\citenamefont {Hagedorn}(1998)}]{Hagedorn:1998}%
  \BibitemOpen
  \bibfield  {author} {\bibinfo {author} {\bibfnamefont {G.~A.}\ \bibnamefont
  {Hagedorn}},\ }\href {https://doi.org/10.1006/aphy.1998.5843} {\bibfield
  {journal} {\bibinfo  {journal} {Ann.\ Phys.\ (NY)}\ }\textbf {\bibinfo
  {volume} {269}},\ \bibinfo {pages} {77} (\bibinfo {year} {1998})}\BibitemShut
  {NoStop}%
\bibitem [{\citenamefont {Ohsawa}\ and\ \citenamefont
  {Leok}(2013)}]{Ohsawa_Leok:2013}%
  \BibitemOpen
  \bibfield  {author} {\bibinfo {author} {\bibfnamefont {T.}~\bibnamefont
  {Ohsawa}}\ and\ \bibinfo {author} {\bibfnamefont {M.}~\bibnamefont {Leok}},\
  }\href {https://doi.org/10.1088/1751-8113/46/40/405201} {\bibfield  {journal}
  {\bibinfo  {journal} {J.~Phys.~A}\ }\textbf {\bibinfo {volume} {46}},\
  \bibinfo {pages} {405201} (\bibinfo {year} {2013})}\BibitemShut {NoStop}%
\bibitem [{\citenamefont {Moghaddasi~Fereidani}\ and\ \citenamefont
  {Van\'i\v{c}ek}(2023)}]{Fereidani_Vanicek:2023}%
  \BibitemOpen
  \bibfield  {author} {\bibinfo {author} {\bibfnamefont {R.}~\bibnamefont
  {Moghaddasi~Fereidani}}\ and\ \bibinfo {author} {\bibfnamefont {J.~J.~L.}\
  \bibnamefont {Van\'i\v{c}ek}},\ }\href {https://doi.org/10.1063/5.0165489}
  {\bibfield  {journal} {\bibinfo  {journal} {J.~Chem.\ Phys.}\ }\textbf
  {\bibinfo {volume} {159}},\ \bibinfo {pages} {094114} (\bibinfo {year}
  {2023})}\BibitemShut {NoStop}%
\bibitem [{\citenamefont {Kl\={e}tnieks}\ and\ \citenamefont
  {Van\'{i}\v{c}ek}(2026)}]{Kletnieks_Vanicek:2023a}%
  \BibitemOpen
  \bibfield  {author} {\bibinfo {author} {\bibfnamefont {E.}~\bibnamefont
  {Kl\={e}tnieks}}\ and\ \bibinfo {author} {\bibfnamefont {J.~J.~L.}\
  \bibnamefont {Van\'{i}\v{c}ek}},\ }\href@noop {} {\enquote {\bibinfo {title}
  {Time-reversible and norm-conserving high-order integrators for the ab initio
  thawed {Gaussian} approximation},}\ } (\bibinfo {year} {2026}),\ \bibinfo
  {note} {not published}\BibitemShut {NoStop}%
\bibitem [{\citenamefont {Moghaddasi~Fereidani}\ and\ \citenamefont
  {Van\'{i}\v{c}ek}(2023)}]{Fereidani_Vanicek:2023a}%
  \BibitemOpen
  \bibfield  {author} {\bibinfo {author} {\bibfnamefont {R.}~\bibnamefont
  {Moghaddasi~Fereidani}}\ and\ \bibinfo {author} {\bibfnamefont {J.~J.~L.}\
  \bibnamefont {Van\'{i}\v{c}ek}},\ }\href {https://doi.org/10.1063/5.0180070}
  {\bibfield  {journal} {\bibinfo  {journal} {J.~Chem.\ Phys.}\ }\textbf
  {\bibinfo {volume} {160}},\ \bibinfo {pages} {044113} (\bibinfo {year}
  {2023})}\BibitemShut {NoStop}%
\bibitem [{\citenamefont {K{\"o}ppel}, \citenamefont {Domcke},\ and\
  \citenamefont {Cederbaum}(1984)}]{Koppel_Cederbaum:1984}%
  \BibitemOpen
  \bibfield  {author} {\bibinfo {author} {\bibfnamefont {H.}~\bibnamefont
  {K{\"o}ppel}}, \bibinfo {author} {\bibfnamefont {W.}~\bibnamefont {Domcke}},\
  and\ \bibinfo {author} {\bibfnamefont {L.~S.}\ \bibnamefont {Cederbaum}},\
  }\href {https://doi.org/10.1002/9780470142813.ch2} {\bibfield  {journal}
  {\bibinfo  {journal} {Adv.\ Chem.\ Phys.}\ }\textbf {\bibinfo {volume}
  {57}},\ \bibinfo {pages} {59} (\bibinfo {year} {1984})}\BibitemShut {NoStop}%
\bibitem [{\citenamefont {Domcke}, \citenamefont {Yarkony},\ and\ \citenamefont
  {K{\"o}ppel}(2004)}]{book_Domcke_Koppel:2004}%
  \BibitemOpen
  \bibfield  {author} {\bibinfo {author} {\bibfnamefont {W.}~\bibnamefont
  {Domcke}}, \bibinfo {author} {\bibfnamefont {D.}~\bibnamefont {Yarkony}},\
  and\ \bibinfo {author} {\bibfnamefont {H.}~\bibnamefont {K{\"o}ppel}},\
  }\href@noop {} {\emph {\bibinfo {title} {Conical intersections: electronic
  structure, dynamics \& spectroscopy}}},\ Vol.~\bibinfo {volume} {15}\
  (\bibinfo  {publisher} {World Scientific},\ \bibinfo {year}
  {2004})\BibitemShut {NoStop}%
\bibitem [{\citenamefont {Begu\v{s}i\'{c}}, \citenamefont {Tapavicza},\ and\
  \citenamefont {Van\'i\v{c}ek}(2022)}]{Begusic_Vanicek:2022}%
  \BibitemOpen
  \bibfield  {author} {\bibinfo {author} {\bibfnamefont {T.}~\bibnamefont
  {Begu\v{s}i\'{c}}}, \bibinfo {author} {\bibfnamefont {E.}~\bibnamefont
  {Tapavicza}},\ and\ \bibinfo {author} {\bibfnamefont {J.}~\bibnamefont
  {Van\'i\v{c}ek}},\ }\href {https://doi.org/10.1021/acs.jctc.2c00030}
  {\bibfield  {journal} {\bibinfo  {journal} {J.~Chem.\ Theory Comput.}\
  }\textbf {\bibinfo {volume} {18}},\ \bibinfo {pages} {3065} (\bibinfo {year}
  {2022})}\BibitemShut {NoStop}%
\bibitem [{\citenamefont {Barbiero}\ and\ \citenamefont
  {Van\'{\i}\v{c}ek}(2026)}]{Barbiero_Vanicek:2026}%
  \BibitemOpen
  \bibfield  {author} {\bibinfo {author} {\bibfnamefont {D.}~\bibnamefont
  {Barbiero}}\ and\ \bibinfo {author} {\bibfnamefont {J.~J.~L.}\ \bibnamefont
  {Van\'{\i}\v{c}ek}},\ }\href {https://doi.org/10.1063/5.0327624} {\bibfield
  {journal} {\bibinfo  {journal} {J.~Chem.\ Phys.}\ }\textbf {\bibinfo {volume}
  {164}},\ \bibinfo {pages} {234114} (\bibinfo {year} {2026})}\BibitemShut
  {NoStop}%
\bibitem [{\citenamefont {Pattanayak}\ and\ \citenamefont
  {Schieve}(1994)}]{Pattanayak_Schieve:1994}%
  \BibitemOpen
  \bibfield  {author} {\bibinfo {author} {\bibfnamefont {A.~K.}\ \bibnamefont
  {Pattanayak}}\ and\ \bibinfo {author} {\bibfnamefont {W.~C.}\ \bibnamefont
  {Schieve}},\ }\href {https://doi.org/10.1103/PhysRevE.50.3601} {\bibfield
  {journal} {\bibinfo  {journal} {Phys.\ Rev.~E}\ }\textbf {\bibinfo {volume}
  {50}},\ \bibinfo {pages} {3601} (\bibinfo {year} {1994})}\BibitemShut
  {NoStop}%
\bibitem [{\citenamefont {Hairer}, \citenamefont {Lubich},\ and\ \citenamefont
  {Wanner}(2006)}]{book_Hairer_Wanner:2006}%
  \BibitemOpen
  \bibfield  {author} {\bibinfo {author} {\bibfnamefont {E.}~\bibnamefont
  {Hairer}}, \bibinfo {author} {\bibfnamefont {C.}~\bibnamefont {Lubich}},\
  and\ \bibinfo {author} {\bibfnamefont {G.}~\bibnamefont {Wanner}},\ }\href
  {http://books.google.ch/books/about/Geometric_Numerical_Integration.html?id=%
T1TaNRLmZv8C&redir_esc=y} {\emph {\bibinfo {title} {Geometric Numerical
  Integration: Structure-Preserving Algorithms for Ordinary Differential
  Equations}}}\ (\bibinfo  {publisher} {Springer Berlin Heidelberg New York},\
  \bibinfo {year} {2006})\BibitemShut {NoStop}%
\bibitem [{\citenamefont {Choi}\ and\ \citenamefont
  {Van\'{i}\v{c}ek}(2021{\natexlab{b}})}]{Choi_Vanicek:2021a}%
  \BibitemOpen
  \bibfield  {author} {\bibinfo {author} {\bibfnamefont {S.}~\bibnamefont
  {Choi}}\ and\ \bibinfo {author} {\bibfnamefont {J.}~\bibnamefont
  {Van\'{i}\v{c}ek}},\ }\href {https://doi.org/10.1063/5.0061878} {\bibfield
  {journal} {\bibinfo  {journal} {J.~Chem.\ Phys.}\ }\textbf {\bibinfo {volume}
  {155}},\ \bibinfo {pages} {124104} (\bibinfo {year}
  {2021}{\natexlab{b}})}\BibitemShut {NoStop}%
\bibitem [{\citenamefont {Petersen}\ and\ \citenamefont
  {Pedersen}(2012)}]{Petersen_Pedersen:2012}%
  \BibitemOpen
  \bibfield  {author} {\bibinfo {author} {\bibfnamefont {K.~B.}\ \bibnamefont
  {Petersen}}\ and\ \bibinfo {author} {\bibfnamefont {M.~S.}\ \bibnamefont
  {Pedersen}},\ }\href {http://www2.imm.dtu.dk/pubdb/p.php?3274} {\enquote
  {\bibinfo {title} {The matrix cookbook},}\ } (\bibinfo {year}
  {2012})\BibitemShut {NoStop}%
\bibitem [{\citenamefont {Van\'{i}\v{c}ek}\ and\ \citenamefont
  {Choi}(2026)}]{Vanicek_Choi:2026}%
  \BibitemOpen
  \bibfield  {author} {\bibinfo {author} {\bibfnamefont {J.~J.~L.}\
  \bibnamefont {Van\'{i}\v{c}ek}}\ and\ \bibinfo {author} {\bibfnamefont
  {S.}~\bibnamefont {Choi}},\ }\href@noop {} {\enquote {\bibinfo {title}
  {{High-order geometric integrators for Ehrenfest dynamics}},}\ } (\bibinfo
  {year} {2026}),\ \bibinfo {note} {not published}\BibitemShut {NoStop}%
\bibitem [{\citenamefont {Runeson}\ and\ \citenamefont
  {Richardson}(2019)}]{Runeson_Richardson:2019}%
  \BibitemOpen
  \bibfield  {author} {\bibinfo {author} {\bibfnamefont {J.~E.}\ \bibnamefont
  {Runeson}}\ and\ \bibinfo {author} {\bibfnamefont {J.~O.}\ \bibnamefont
  {Richardson}},\ }\href {https://doi.org/10.1063/1.5100506} {\bibfield
  {journal} {\bibinfo  {journal} {J.~Chem.\ Phys.}\ }\textbf {\bibinfo {volume}
  {151}},\ \bibinfo {pages} {044119} (\bibinfo {year} {2019})}\BibitemShut
  {NoStop}%
\end{thebibliography}

%
\end{document}